\documentstyle[12pt]{article}

\input epsf

\begin{document}
\baselineskip=20.5pt

\def\beqra{\begin{eqnarray}} \def\eeqra{\end{eqnarray}}
\def\beqast{\begin{eqnarray*}}
\def\eeqast{\end{eqnarray*}}
\def\beq{\begin{equation}}      \def\eeq{\end{equation}}
\def\be{\begin{enumerate}}   \def\ee{\end{enumerate}}

\def\fnote#1#2{\begingroup\def\thefootnote{#1}\footnote{
#2}
\addtocounter
{footnote}{-1}\endgroup}

\def\technion#1#2{\hfill{TECHNION-{#1}-{#2}}}

\def\gam{\gamma}
\def\Gam{\Gamma}
\def\la{\lambda}
\def\eps{\epsilon}
\def\La{\Lambda}
\def\si{\sigma}
\def\Si{\Sigma}
\def\al{\alpha}
\def\Tha{\Theta}
\def\tha{\theta}
\def\vphi{\varphi}
\def\del{\delta}
\def\Del{\Delta}
\def\ab{\alpha\beta}
\def\om{\omega}
\def\Om{\Omega}
\def\mn{\mu\nu}
\def\mun{^{\mu}{}_{\nu}}
\def\kap{\kappa}
\def\rsi{\rho\sigma}
\def\beal{\beta\alpha}

\def\til{\tilde}
\def\rta{\rightarrow}
\def\eqv{\equiv}
\def\nab{\nabla}
\def\pa{\partial}
\def\sit{\tilde\sigma}
\def\ul{\underline}
\def\indt{\parindent2.5em}
\def\nd{\noindent}

\def\rsi{\rho\sigma}
\def\beal{\beta\alpha}

\def\caa{{\cal A}}
\def\cb{{\cal B}}
\def\cac{{\cal C}}
\def\cd{{\cal D}}
\def\ce{{\cal E}}
\def\cf{{\cal F}}
\def\cg{{\cal G}}
\def\cah{{\cal H}}
\def\ci{{\cal I}}
\def\cj{{\cal{J}}}
\def\ck{{\cal K}}
\def\cl{{\cal L}}
\def\cm{{\cal M}}
\def\cn{{\cal N}}
\def\cO{{\cal O}}
\def\cp{{\cal P}}
\def\car{{\cal R}}
\def\cs{{\cal S}}
\def\ct{{\cal{T}}}
\def\cu{{\cal{U}}}
\def\cv{{\cal{V}}}
\def\cw{{\cal{W}}}
\def\cx{{\cal{X}}}
\def\cy{{\cal{Y}}}
\def\cz{{\cal{Z}}}

\def\raisenot{\raise .5mm\hbox{/}}
\def\nota{\ \hbox{{$a$}\kern-.49em\hbox{/}}}
\def\notA{\hbox{{$A$}\kern-.54em\hbox{\raisenot}}}
\def\notb{\ \hbox{{$b$}\kern-.47em\hbox{/}}}
\def\notB{\ \hbox{{$B$}\kern-.60em\hbox{\raisenot}}}
\def\notc{\ \hbox{{$c$}\kern-.45em\hbox{/}}}
\def\notd{\ \hbox{{$d$}\kern-.53em\hbox{/}}}
\def\notbd{\ \hbox{{$D$}\kern-.61em\hbox{\raisenot}}} 
\def\note{\ \hbox{{$e$}\kern-.47em\hbox{/}}}
\def\notk{\ \hbox{{$k$}\kern-.51em\hbox{/}}}
\def\notp{\ \hbox{{$p$}\kern-.43em\hbox{/}}}
\def\noto{\ \hbox{{$o$}\kern-.43em\hbox{/}}}
\def\notq{\ \hbox{{$q$}\kern-.47em\hbox{/}}}
\def\notW{\ \hbox{{$W$}\kern-.75em\hbox{\raisenot}}}
\def\notz{\ \hbox{{$Z$}\kern-.61em\hbox{\raisenot}}}
\def\notpa{\hbox{{$\partial$}\kern-.54em\hbox{\raisenot}}}

\def\fo{\hbox{{1}\kern-.25em\hbox{l}}}  
\def\rf#1{$^{#1}$}
\def\bx{\Box}
\def\tr{{\rm Tr}}
\def\rmtr{{\rm tr}}
\def\dgg{\dagger}

\def\lag{\langle}
\def\rag{\rangle}
\def\bmid{\big|}

\def\vlap{\overrightarrow{\La p}} 
\def\lrta{\longrightarrow}
\def\lrar{\raisebox{.8ex}{$\longrightarrow$}}
\def\rlarw{\longleftarrow\!\!\!\!\!\!\!\!\!\!\!\lrar}

\def\llra{\relbar\joinrel\longrightarrow}     
\def\mapright#1{\smash{\mathop{\llra}\limits_{#1}}}
\def\mapup#1{\smash{\mathop{\llra}\limits^{#1}}}
\def\asymptotic{{_{\stackrel{\displaystyle\longrightarrow}
{x\rightarrow\pm\infty}}\,\, }} 
\def\asymptext{\raisebox{.6ex}{${_{\stackrel{\displaystyle\longrightarrow}
{x\rightarrow\pm\infty}}\,\, }$}} 

\def\7#1#2{\mathop{\null#2}\limits^{#1}}   
\def\5#1#2{\mathop{\null#2}\limits_{#1}}   
\def\too#1{\stackrel{#1}{\to}}
\def\tooo#1{\stackrel{#1}{\longleftarrow}}
\def\nout{{\rm in \atop out}}

\def\one{\raisebox{.5ex}{1}}
\def\BM#1{\mbox{\boldmath{$#1$}}}

\def\ltsim{\matrix{<\cr\noalign{\vskip-7pt}\sim\cr}}
\def\gtsim{\matrix{>\cr\noalign{\vskip-7pt}\sim\cr}}
\def\haf{\frac{1}{2}}


\def\place#1#2#3{\vbox to0pt{\kern-\parskip\kern-7pt
                             \kern-#2truein\hbox{\kern#1truein #3}
                             \vss}\nointerlineskip}

\def\illustration #1 by #2 (#3){\vbox to #2{\hrule width #1
height 0pt
depth
0pt
                                       \vfill\special{illustration #3}}}

\def\scaledillustration #1 by #2 (#3 scaled #4){{\dimen0=#1
\dimen1=#2
           \divide\dimen0 by 1000 \multiply\dimen0 by #4
            \divide\dimen1 by 1000 \multiply\dimen1 by #4
            \illustration \dimen0 by \dimen1 (#3 scaled #4)}}

\def\ON{{\cal O}(N)}
\def\UN{{\cal U}(N)}
\def\bdPh{\mbox{\boldmath{$\dot{\!\Phi}$}}}
\def\bPh{\mbox{\boldmath{$\Phi$}}}
\def\bPhs{\bPh^2}
\def\sef{S_{eff}}
\def\sigx{\sigma(x)}
\def\pix{\pi(x)}
\def\bph{\mbox{\boldmath{$\phi$}}}
\def\bphs{\bph^2}
\def\ex{\BM{x}}
\def\exs{\ex^2}
\def\xdot{\dot{\!\ex}}
\def\y{\BM{y}}
\def\ys{\y^2}
\def\ydot{\dot{\!\y}}
\def\pat{\pa_t}
\def\pax{\pa_x}

\renewcommand{\theequation}{\arabic{equation}}



\vspace*{.2in}
\begin{center}
\large{\bf Casimir Effect: The Classical Limit}\\
\vspace{26pt}
{J. Feinberg$^{a,b}$, A. Mann$^{b}$  \& M. Revzen$^{b}$\fnote{}{{\it e-mail: joshua, ady, revzen@physics.technion.ac.il}}}
\end{center}

\vskip 2mm
\begin{center}$^{a)}$\fnote{*}{permanent address} Department of Physics, \\
Oranim-University of Haifa, Tivon 36006, Israel\\
$^{b)}$~{Department of Physics,}\\
{Technion - Israel Institute of Technology, Haifa 32000 Israel}
\vspace{.6cm}                                             
\end{center}

\begin{minipage}{6.3in}
{\abstract~~~~~ 
We analyze the high temperature (or classical) limit of the Casimir 
effect. A useful quantity which arises naturally in our discussion is 
the ``relative Casimir energy", which we define for a configuration of 
disjoint conducting boundaries of arbitrary shapes, as the difference 
of Casimir energies between the given configuration and a configuration 
with the same boundaries infinitely far apart. Using path integration 
techniques, we show that the relative Casimir energy vanishes exponentially 
fast in temperature. This is consistent with a simple physical argument 
based on Kirchhoff's law. As a result the ``relative Casimir 
entropy", which we define in an obviously analogous manner, 
tends, in the classical limit, to a finite asymptotic value which depends 
only on the geometry of the boundaries. Thus the Casimir force 
between disjoint pieces of the boundary, in the classical limit, is 
entropy driven and is governed by a dimensionless number characterizing the 
geometry of the cavity. Contributions to the 
Casimir thermodynamical quantities due to each individual connected 
component of the boundary exhibit logarithmic deviations in temperature 
from the behavior just described. These logarithmic deviations seem to arise 
due to our difficulty to separate the Casimir energy (and the other 
thermodynamical quantities) from the ``electromagnetic'' self-energy of 
each of the connected components of the boundary in a well defined manner. 
Our approach to the Casimir effect is not to impose sharp boundary conditions 
on the fluctuating field, but rather take into consideration its 
interaction with the plasma of ``charge carriers'' in the boundary,
with the plasma frequency playing the role of a physical 
UV cutoff. This also allows us to analyze deviations from
a perfect conductor behavior.} 
\end{minipage}

\vspace{28pt}
\vfill
\pagebreak

\setcounter{page}{1}

\section{Introduction}     


There has been considerable interest in the Casimir effect \cite{casimir} lately\cite{cas50}. 
Recent discussions of this effect and its variants range over several areas of 
physics \cite{recently, lambib, kardar, kardar1, nussinov1, nussinov2, amj, opher, ford}. For recent books and reviews see 
\cite{milonni,trunov,levin, krech, miltrev}. For earlier accounts see \cite{greiner, zuber}.

The experimental status of the effect was tenuous until 1997 when vindication of the theoretical prediction to an 
accuracy of a few percent was reported \cite{prl}.

Over the years, various methods have been applied to analyze the Casimir effect. To name a few, we mention 
the ``mode summation technique" \cite{milonni, trunov, greiner, zuber} (which was the method 
originally used by Casimir), analyses of Green's functions and of the energy-momentum tensor in the presence of 
boundaries \cite {fierz, brown, boyer, schwinger, mrssphere, miltonsphere, takagi} and zeta function 
techniques \cite{zeta, elizeta}.  These techniques possess flexibility which permits, among other things, 
the handling of temperature dependence (in vacuum and in dielectrics) \cite{fierz, brown, takagi, lifshitz, mrs, wolfram, baliandup, 
bordagt}, of fields with finite mass \cite{wolfram, hays}, 
various geometrical setups \cite{kardar, kardar1, nussinov1, nussinov2, boyer, mrssphere, miltonsphere, baliandup, balian, 
brevikold, candelasphere}, and a better insight into the Casimir 
force (macroscopic) in relation to microscopic ``long range forces'' \cite {levin}.

Finally, we mention here the path integral approach 
\cite{kardar, kardar1, nussinov1, nussinov2, bordag} which seems to be well suited to handle 
systems with boundaries of arbitrary shape as well as systems with moving boundaries \cite{dynamic}.

An important aspect of the Casimir effect is its behavior at finite temperatures.
Finite temperature studies were carried out by several authors, e.g., 
\cite{fierz, brown, takagi, lifshitz, mrs, wolfram, bordagt}. In particular, we mention in this 
context the thorough analysis by Balian and Duplantier \cite{baliandup}, who used 
multiple scattering expansions to study the Casimir effect at finite temperature in the presence of conducting 
boundaries of arbitrary smooth shapes. One of the several issues studied in \cite{baliandup} 
was the classical (i.e., high temperature) limit of the effect (see Section 6 in \cite{baliandup}).

In this paper we address the classical (or high temperature) limit of the 
Casimir effect due to boundaries of arbitrary shapes using path integration.
Our results are in agreement with those of \cite{baliandup}. 
The classical limit is attained upon subtraction of the ``self-energies'' 
of the boundaries. The need of this subtraction for obtaining the classical 
limit of the Casimir effect leasds us to define in (Section 2 of ) this 
paper what we call the ``relative Casimir free-energy", the ``relative 
Casimir energy" and the ``relative Casimir entropy". In fact, based on 
Kirchhoff's law, we can explain the classical limit of the Casimir 
effect intuitively in terms of these newly defined relative quantities.\\

To get oriented, let us briefly review (hopefully in a self contained manner) the standard Casimir 
problem, viz., radiation field confined between two parallel conducting plates, at zero and 
at finite temperature. We will write down explicit expressions for the Casimir energy, free energy and 
entropy. These are displayed in Fig. 1, which gives the behavior of these quantities
for all temperatures. At the end of this section we will discuss the classical limit of the Casimir 
effect in this simple geometry. (Readers familiar with these standard aspects of the Casimir effect may skip directly to the next section, but are urged to take a look in Fig. 1 on the way.)
In Appendix C we present an alternative treatment of this problem, based on the path integral formalism, following 
closely \cite{kardar, nussinov1, nussinov2}.

For simplicity, we will employ, for the rest of 
this section, the so called ``mode summation technique". (For more details see, e.g., 
\cite{opher, milonni, trunov, greiner, zuber}. Here we follow closely \cite{opher}.) We consider the 
radiation field (in $3+1$ space-time dimensions) confined between two conducting parallel plates. 
(In later sections, where we discuss the high temperature limit of the Casimir effect for 
cavities of arbitrary shapes, we will limit ourselves to massless scalar fields\footnote{This 
simplification does not spoil the agreement between our results and the results in Section 6 of 
\cite{baliandup} which discusses the electromagnetic case.} for simplicity.)
The sizes of the plates' edges are L. The parallel plates coincide with the planes at $z = 0$ and $z=d$.
We are interested of course in the case where $L >>d$, and in the end will take the limit $L \rightarrow \infty$.

The energy tied down in zero point fluctuations in the 
mode $k$ is, in obvious notation ($k$ includes the polarization and the zero in the argument relates to 
$T = 0$),
\begin{equation}\label{zpe}
E_k(0) = \frac{1}{2}\hbar \omega_k = \frac{1}{2} \hbar c|k|\,.
\end{equation}
  
The total energy density of the EM field with the conducting plates as
boundaries is ($k_\perp$ is the magnitude of the wave vector perpendicular to the $z$ axis, i.e., parallel to 
the plates \cite{zuber})
\begin{equation}
\ce (d, T=0) = \frac{\sum_{k}E_{k}}{L^{2}d} = 
\frac{\hbar c}{2\pi d}\int^{\infty}_0 k_\perp dk_\perp\left[ \frac{k_\perp}{2} + 
\sum_{m=1}^{\infty}|k_m|\right]
\end{equation}
$$ k_\perp^2 = k_{x}^2 + k_{y}^2; \;\; k_{m}^2 = k_\perp^2 +
 \frac{m^2 \pi^2}{d^2}; \;\; m = 1,2,...$$

The energy density, in dimensionless units, $\varepsilon(0)$, is given by
\cite{opher}
\begin{equation}\label{casenergy0}
\ce (d,0) = {\pi^2 \hbar c\over 2 d^4}\varepsilon(0)
\equiv D\varepsilon(0);
\end{equation}
\begin{equation}
\varepsilon(0) = \int^{\infty}_0 xdx\ \left[\frac{x}{2} + \sum_{m=1}^{\infty}
\sqrt{x^2 + m^2}\right].
\end{equation}
Equation (\ref{casenergy0}) defines the energy density scale D. 
The energy density due to vacuum fluctuations of the radiation field in an 
arbitrarily large volume $V = L^3$ - which serves as the 
reference (unconstrained) system -  is
\begin{equation}
\ce (\infty,0) = \frac{1}{V}\sum_k \frac{1}{2}\hbar 
\omega_k = D \varepsilon_\infty (0),
\end{equation}
$$\varepsilon_\infty (0) = \int\limits_0^\infty xdx\ \int\limits_0^\infty dm\sqrt{x^2 + m^2}.$$
The Casimir energy density, at $T=0$, in dimensionless units, is given by
\begin{equation}\label{casendef}
\varepsilon_{c}(0) = \varepsilon(0) - \varepsilon_\infty (0).
\end{equation}                                                              
Both $\varepsilon(0)$ and $\varepsilon_\infty (0)$ diverge. $\varepsilon_c$ 
is commonly \cite{casimir, milonni, greiner} evaluated by a  physically 
justifiable regularization
technique. Thus a wave vector dependent function, $r(k/k_c)$, is introduced
into the above integrals 
such that for $k >> k_c$\ $r \rightarrow 0$ while $r \rightarrow 1$ for 
$k << k_c$, thereby rendering the integrals convergent. A simple choice is 
($\alpha = 1/k_c$), 
\beq\label{expreg}
r(\alpha  k) = \exp \left[-\alpha k \right]\,.
\eeq
$\alpha$ is allowed to go to zero at the end of the calculations - for all 
the terms together - thus the sum is ``regularized''. The details of the 
calculations will not be given here (cf. \cite {opher, greiner, zuber}.) In Appendix C, where we present
an alternative analysis of this problem, we employ a different regularization scheme for the Casimir energy,
which seems to us more natural physically. 

The result for our case is
\begin{equation}\label{casen}
\varepsilon_{c} (0) = -\frac{4}{(2\pi)^4} \zeta(4) = -{1\over 360}\,,
\end{equation}
or, in physical units,
\beq\label{casenphys}
\ce_c (d, 0) = D\varepsilon_c (0) = -{\pi^2\hbar c\over 720 d^4}\,.
\eeq
The force $f_c(d, 0)$ per unit area between the plates is now calculable from the 
regularized Casimir energy (\ref{casenphys}),  yielding the well-known Casimir force \cite{casimir}
\begin{equation}\label{casimirforce}
f_c(d, 0) = - {1\over L^2} \frac{\pa (L^2 d \ce_c)}{\pa d} = - \frac{\pi^2\hbar c}{240 d^4}\,.
\end{equation}

The finite temperature problem is quite similar. Now the zero point energy is
supplemented by the thermal energy. Thus the (average) energy tied down in 
the $k$-th mode is
\begin{equation}\label{aenergy}
E_k (T) = (1/2)\hbar \omega_k + \frac{\hbar \omega_k}{\exp(\beta \hbar
 \omega_k) - 1}
 = \frac{\hbar \omega_k}{2}{\rm coth}\left[\frac{\beta \hbar \omega_k}{2}\right].
\end{equation}
Correspondingly, the energy density is 
\begin{equation}\label{et}
\ce (d,T) = \frac{\hbar c}{2\pi d}\int k_\perp dk_\perp
\left[\frac{|k_\perp|}{2}{\rm coth}(\frac{\beta \hbar \omega(k_\perp)}{2}) 
+ \sum_{m=1}^{\infty}|k_m|{\rm coth}(\frac{\beta \hbar \omega(k_m)}{2})\right].
\end{equation}
Returning to our dimensionless units we may write the energy density as a sum 
of a zero temperature part plus a temperature dependent part:
\begin{equation}\label{edt}
\varepsilon(d,T) = \varepsilon(0) + u(d,T),
\end{equation}
\begin{equation}\label{usum}
u(d,T) = \frac{f(0)}{2} + \sum_{m=1}^{\infty}f(m),
\end{equation}
$$ f(m) = \int\limits_{m}^{\infty}dy\ {y}^2 n(y,T),$$
\begin{equation}\label{nyt}
n(y,T) = \frac{2}{\exp(\frac{T_c}{T}y) - 1}\,, 
\end{equation}
with 
\begin{equation}\label{tc}
k{_B}T{_c} = \hbar c\frac{\pi}{d}\,.
\end{equation}
Observe that for dimensional reasons, $\varepsilon(d,T), u(d,T)$ and $n(y,T)$ (as well as all the other 
thermodynamical quantities which we will define below) depend on $d$ and $T$
through the combination $ Td \propto  T/T_c $ only. Thus, in particular, $\varepsilon (d, T=0) = \varepsilon (0)$ is 
independent of $d$.

We see from (\ref{nyt}) that $T_c$ plays the role of a characteristic (or lower cutoff) temperature of the
geometrical setting, separating high and 
low temperatures.
The corresponding expression for the unconstrained system is
\begin{equation}
\varepsilon(\infty,T) = \varepsilon(0) +u(\infty,T),
\end{equation}
with $$ u(\infty,T) = \int\limits_{0}^{\infty}dm\ f(m)\,.$$
Evaluating the sum in (\ref{usum}) via the Poisson summation formula \cite{opher, greiner} gives ($\mu = 2\pi m$)
\begin{equation}\label{ect} 
\varepsilon_c(t) \equiv \varepsilon(d,T) - \varepsilon(\infty,T) = 
-4t^3\sum_{m=1}^{\infty}\frac{1}{\mu}{\rm coth}(t\mu){\rm csch}^2(t
\mu),\; \; \; t = \pi \frac{T}{T_c}.
\end{equation}

As a consistency check we observe that for $t << 1, \;\;\varepsilon_c \rightarrow \varepsilon_c(0)$, the 
expression in (\ref{casen}). The finite temperature Casimir energy 
$\varepsilon_c(t)$, i.e., the difference between the expectation values of the Hamiltonians of the 
constrained and unconstrained systems, is displayed in Fig.1.

The Casimir free energy is calculated as follows:
The partition function (for one mode) is 
\begin{equation}\label{partk}
Z_k = \sum_{n=1}^{\infty}\exp\left[-\beta\hbar \omega_k(n +1/2)\right] = 
\frac{\exp\left[-\beta\hbar \omega_k/2\right]}{1 - \exp\left[-\beta\hbar 
\omega_k\right]},
\end{equation}
and thus the free energy for the k-th mode is 
\begin{equation}\label{freek}
F_k = - k_B T \,{\rm log}\, Z_k = \frac{1}{2}\hbar \omega_k + k_B T \log 
(1 -\ \exp\left[-\beta\hbar \omega_k\right])\,.
\end{equation}
Hence the expression for the free energy density of the constrained system is
\begin{equation}\label{freet}
\cf (d,T) = \ce (d,0) + k_{B}T\frac{\hbar c}{2\pi 
d}\int k_\perp dk_\perp\left[ \log (1 -e^{-\beta \hbar c k_\perp}) + 
2\sum_{m=1}^{\infty} \log (1 - e^{-\beta \hbar c k_{m}})\right].
\end{equation}
A  corresponding equation holds for the unconstrained 
system, thereby leading to the 
dimensionless expression for the Casimir free energy density, $ \phi_c $,
 given by
\begin{equation}\label{freecphys}
\cf_c(T) = D \phi_c(t),
\end{equation}
where D was defined in (\ref{casenergy0}), and 
\begin{equation}\label{freec}
\phi_c(t) = -2t\sum_{m=1}^{\infty} {\frac{1}{\mu^3}}\left[{\rm coth}
(t\mu) +
 (t\mu){\rm csch}^2(t\mu)\right] .
\end{equation}
The Casimir free energy,
$\phi_c(t)$, is displayed in Fig. 1. As a check, note that for $t << 1$, $\phi_c(t) \rightarrow 
\varepsilon_c(0)$, as it should.

At this juncture it is natural to consider Casimir's entropy 
\cite{opher, brown}. This entropy, $\sigma_c(t)$, is defined in our 
dimensionless units by
\begin{equation}\label{expfall}
\phi_c(t) = \varepsilon_c(t) - t\sigma_c(t).
\end{equation}
$ -\sigma_c(t)$ is displayed in Fig. 1.

At high 
temperatures, $t >> 1$, $\varepsilon_c(t)$ 
falls off exponentially with $t$, $\sigma_c(t)$ approaches a constant value  
(independent of t and, of course, 
of $\hbar$) while $\phi_c(t)$ becomes proportional to $t$. (An explicit calculation of this
behavior is deferred to Appendix A.) Reexpressing 
these results in standard physical dimensions we get, at this limit ($t >> 1$)
\begin{equation}\label{classicallimit}
\ce_c \rightarrow 0, \; \; \; \cf_c \rightarrow -T\cs_c, \; \; \; \cs_c \rightarrow 
\frac{\zeta(3)}{2^3 \pi}(\frac{L^2}{d^2}).
\end{equation}
Thus the entropy is proportional to the area of the plates scaled by $d^2$. 
In view of this fact we conclude that in the classical limit (high temperatures) the Casimir force 
is purely entropic.\\{}\\ 
\newpage
\vspace*{-6cm}
\begin{figure}[htbp]
\epsfxsize=0.8\textwidth
\centerline{\epsffile{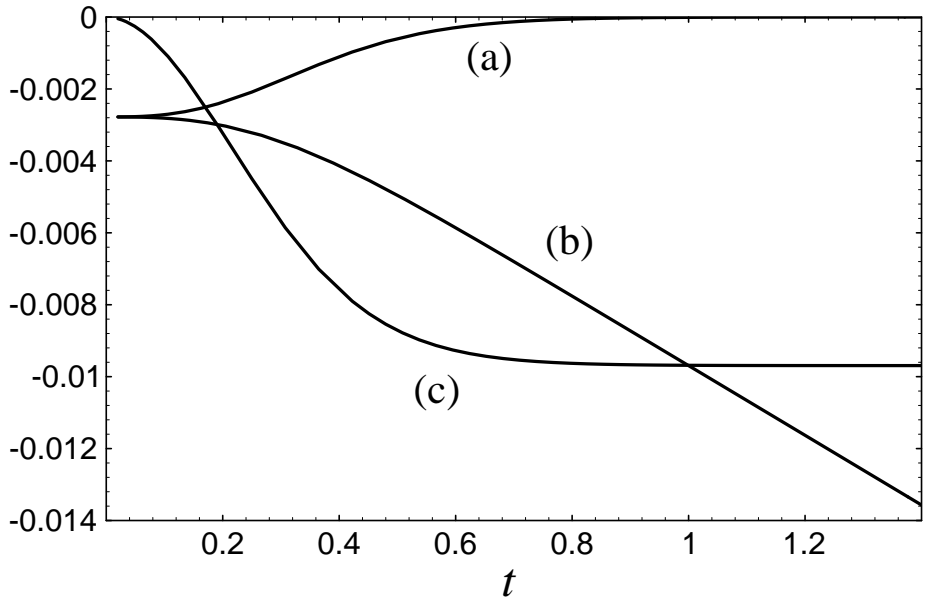}}
\vspace*{-4.5cm}
\caption{Casimir's Energy $\varepsilon_c$ (a), 
Free Energy $\phi_c$ (b), and 
Entropy $-\sigma_c$ (c) as a function of temperature, $t$. Note that graphs (a) and (b) coincide at energy 
$-0.0028\simeq -1/360$, consistent with (\ref{casen}).}
\end{figure}

It is clear from Fig. 1 that the Casimir energy vanishes at high temperatures (and according to 
(\ref{ect}) it vanishes exponentially fast). This, we contend, 
characterizes the classical limit where the energy equipartition ``theorem" and the validity of the one to one 
correspondence between the states of the constrained system and the states of the unconstrained system imply 
this result. This is discussed in more detail in Section 2.

This concludes the general introductory review.
The rest of the paper is organized as follows. Section 2 includes the central point 
of the paper, viz., an exposition of the classical limit and a study of its implications.
It is based on the argument that the known \cite{amj} one to one correspondence between states of 
constrained systems and essentially ``free'' ones implies the vanishing of 
the Casimir energy. This is a somewhat subtle issue. To be precise, what vanishes in the 
classical limit is what we define as the {\em relative} Casimir energy, namely, the Casimir energy 
associated with a given {\em disconnected} set of boundaries, 
from which we subtract the Casimir energy for a system in which all these boundaries are infinitely 
separated from one another. In particular, this statement does {\em not} hold for a boundary made of 
a single connected piece. In contrast, the relative Casimir entropy does not vanish in the classical limit.  
We also summarize in Section 2 the high temperature results of Section 6 of \cite{baliandup}, which support these 
assertions about the classical limit for systems constrained by conducting boundaries of arbitrary smooth shapes.
In Section 3 we study, in the path integral formalism, the Casimir effect associated with a massless 
scalar field $\phi(x)$ (as a simple analog of the electromagnetic field), 
confined in a static reflecting cavity of arbitrary shape. Section 3 is divided into several subsections. 
In Section 3.1 we study the zero temperature effect. First, we review the formalism of 
\cite{kardar, nussinov1, nussinov2} assuming sharp boundary conditions on all frequency modes (Section 3.1.1). 
These boundary conditions on the field modes are of course an  approximate macroscopic manifestation of the 
interaction of the ``plasma'' of ``charge carriers'' that ``live'' in the conducting boundary, with the field $\phi$, in 
the limit that the boundary is a perfect conductor.  Thus, in Section 3.1.2 we adopt a more realistic 
approach and take into account the ``susceptibility" of the boundary medium, with its plasma frequency $\om_p$
playing the role of a physical UV cutoff. In the limit of infinite $\om_p$ we recover, of course, the 
effect of sharp boundary conditions. However, by keeping $\om_p$ large but finite, we can also 
calculate corrections to the Casimir effect due to deviations from perfect conductor behavior. In Section 3.2 
we discuss the finite temperature case. The relevant temperature range is $k_BT << \hbar\om_p$. 
In Section 3.2.1 we study the classical limit $k_BT_c < k_BT << \hbar\om_p$, where
$T_c$ is a lower (IR) cutoff temperature analogous to (\ref{tc}). For this model, we are able to verify the 
arguments of Section 2, that the classical limit for this system results in a vanishing  
(relative) Casimir energy and a geometry dependent (relative) Casimir entropy. This we can do only 
when the boundaries are made of disjoint pieces. We find that contributions to the Casimir free energy and entropy due to each connected 
component of the boundary individually exhibit logarithmic deviations in temperature from the behavior 
just described. In the conclusions which are given in Section 4 we review the argument (cf. \cite{milonni}) for 
the existence of zero point fluctuations as viewed from our vantage point.

Some technical details are relegated to the appendices. In Appendix A we provide an explicit calculation of the 
exponential fall-off of $\varepsilon_c(t)$ seen in Fig.1 (which we mentioned following Eq. (\ref{expfall})). 
We show that a naive high temperature expansion for the Casimir energy in powers of $1/T$ does not have any 
corrections to the classical value. In Appendix B we verify the consistency of 
$\om_p(T=0)$ as a UV cutoff at high temperatures. Appendix C provides a derivation of the Casimir effect for the 
canonical parallel plate geometry via the path integral formalism of Section 3. We obtain all the results of Section 1
at zero and finite temperature, as well as the classical limit. Appendix D contains some technical details concerning
bounds on terms in a certain expansion we carry out in Section 3.

\newpage

\section{ The Classical Limit - Discussion}

An intuitive understanding of the vanishing of the Casimir energy of the parallel plate geometry
for $T > T_c$ ( cf. Fig. 1) may be gained via the following reasoning.
Loosely speaking \cite{amj} the total number of normal modes (per unit volume) 
in our confined system is unchanged upon changing $d$ (in the 
case under study in Section 1).\footnote{The calculations of Appendix A constitute an explicit 
`proof' of this claim (within the regularization scheme employed there).}
Thus moving the walls adiabatically leads to shifts
in the levels - not to appearance (or disappearance) of levels (modes), 
and there is an (intuitive) one to one correspondence 
between the levels regardless of the size of $d$.   
This, coupled with the classically valid equipartitioning of energy,
which of course means that each harmonic oscillator-like mode holds $k_BT$ 
amount of energy, implies that the (bulk) energy density
(in this classical limit) is unchanged upon changing $d$. 

Thus, in the classical limit, ${\it defined}$ by the validity of the Rayleigh Jeans (RJ) 
law or the equipartioning of the energy \cite{revzen}, the difference
in the energy density between the constrained ($d << L$) and the 
unconstrained ($d \sim L$) cases, namely, the Casimir energy, is nil. 

Thus we have, in conformity with the result of the exact calculations as
exhibited in Fig. 1, that,
\begin{equation}\label{kirchofff}
\lim_{\frac{T}{T_c} \rightarrow \infty} \varepsilon_c(d,T) = 0.
\end{equation}  
These considerations are readily generalizable to systems made of several disjoint conducting 
boundaries of arbitrary shapes (the parallel plate geometry being an example of such a system). 
In the next section we will prove, by an explicit calculation, that the relative Casimir energy 
$\ce_c^{rel}$, which we define as the difference (in obvious notation)
\beq\label{relenergy}
\ce_c^{rel} (T) = \ce_c (T) - \ce_c^\infty (T)
\eeq
between the Casimir energy associated with a given {\em disconnected} set of boundaries and the Casimir energy for a system in which all these boundaries are infinitely separated from one another (but 
are otherwise unchanged), vanishes in the classical limit
\begin{equation}\label{kirchoff}
\lim_{\frac{T}{T_c} \rightarrow \infty} \ce_c^{rel} (T) = 0\,,
\end{equation}  
where $T_c$ is a characteristic temperature of the constrained system, analogous to (\ref{tc}). (We tacitly 
assume that $T_c$ is the maximal parameter of this kind, if several such parameters are needed to describe 
the constrained system.)

In discussions of black body radiation, Kirchhoff's law (that the ratio of
emissivity to absorptivity of all bodies is a universal function of the 
wavelengh $\lambda$ and the temperature $T$) is used \cite{milonni}
to infer that $U$, the total electromagnetic energy density in a cavity at 
thermal equilibrium, is a function of 
$T$ only, 
\begin{equation}\label{kirchoff100}
U = U(T)\,.
\end{equation}
In particular, it means that $U(T)$ is independent of the shape of the cavity.
It is clear that (\ref{kirchoff100}) is true only for $T > T_c$. (E.g., at $T = 0$ it is, 
strictly speaking, never true.) For this reason we refer to (\ref{kirchoff}) as ``Kirchhoff's theorem''.\footnote{That $U(T)$ depends on the shape of the 
cavity at low temperature, which thus sets limitations on the validity of Kirchhoff's law, is of course well known. See, e.g., the comment made on p. 78 
in \cite{pippard}.}

Coming back to the parallel plate geometry, it follows from (\ref{kirchoff}), via (\ref{classicallimit}), that the Casimir force 
$f_c = -\pa\cf_c/\pa d$ between the plates is purely entropic in the classical limit.  Furthermore, in this 
limit, the Casimir entropy $\cs_c $ (defined as the difference in entropies between the constrained and 
free systems) is independent of temperature and is, hence, purely geometric \cite{opher}.  

More generally, as a result of ``Kirchhoff's theorem'' (\ref{kirchoff}), the Casimir forces between 
{\em disconnected} boundaries become purely entropic in the classical limit. 
Consequently, the relative Casimir entopy $\cs_c^{rel} (T)$, defined (in analogy with the parallel 
plate case) as the difference 
\beq\label{relentropy}
\cs_c^{rel} (T)= \cs_c (T)- \cs_c^\infty (T)\,,
\eeq
between the Casimir entropy $\cs_c$ associated with a given 
disconnected set of boundaries, and the Casimir entropy $\cs_c^\infty$ of a 
system in which all these boundaries are infinitely separated from one another (but are otherwise unchanged), 
becomes, in the classical limit, independent of temperature and hence, purely geometric. 

This is consistent, of course, with physical intuition, 
because at temperatures $T>>T_c$ the dominant contribution to entropy is 
from short wavelength 
modes ($\lambda \sim  1/T$). These are insensitive to the boundaries when the 
dimension of the cavity much exceeds these wavelengths \cite{peierles}. 
This dominant contribution to the total entropy cancels out when one calculates the Casimir entropy, namely, the difference in entropies between the unconstrained and constrained systems.  The leading contribution to the 
Casimir entropy, then, is essentially 
determined by the long wavelengths which are geometry dependent, i.e.,  which relate to the 
``shape'' of the cavity and are independent of temperature.

Let us formalize this argument somewhat further:
The entropy of a non-interacting Bose gas, such as the gas of photons confined in the cavity, is  
\beq\label{entropy}
S (T) = {1\over T}\int\limits_0^\infty d\om\, \rho (\om) \left[ {\hbar\om\over e^{\beta\hbar\om} -1} 
-k_BT \log\,\left(1-e^{-\beta\hbar\om}\right)\right]\,,
\eeq
where $\rho(\om)$ is the (extensive) density of modes. Thus, for photons 
(inside a very large volume $V$) $\rho_{free}(\om) = V\om^2/\pi^2c^3$, whereas for radiation 
modes between the parallel plates of the previous section\cite{amj} 
$$\rho (\om) = (c\pi/\om d) \rho_{free} (\om)\left[ 1/2 + \sum_{n=1}^\infty \theta ((\om d/c\pi) -n)
\right]\,,$$ and so on. 

Based on (\ref{entropy}) we can express the Casimir entropy as
\beq\label{casimirentropy}
\cs_c (T)= {1\over T}\int\limits_0^\infty d\om\, \left(\rho (\om) - \rho_{free} (\om)\right)
 \left[ {\hbar\om\over e^{\beta\hbar\om} -1} 
-k_BT \log\,\left(1-e^{-\beta\hbar\om}\right)\right]\,. 
\eeq
Clearly, the density of modes $\rho (\om)$ in the cavity differs significantly from the density of modes 
$\rho_{free} (\om)$ of the free system only at long wavelengths, i.e., they differ significantly only for 
frequencies below some IR cutoff frequency $\om_c$ which is determined by the shape of the cavity. Thus, the 
spectral integral in (\ref{casimirentropy}) is dominated by $\om<\om_c$. This allows us to define the high 
temperature (classical) limit as such that $\beta\hbar\om<\beta\hbar\om_c \equiv T_c/T <<1$. In this limit 
(and due to the fact that $\rho-\rho_{free}$ cuts the integral in (\ref{casimirentropy}) above $\om_c$) we 
see that  
\beq\label{classentropy1}
\lim_{\frac{T}{T_c} >> 1} \cs_c = k_B\,\int\limits_0^\infty d\om\, \left(\rho (\om) - \rho_{free} (\om)\right)
 \left[ 1 - \log\,(\beta\hbar\om)\right]\,. 
\eeq

We will treat 
\beq\label{casdensity}
\delta\rho (\om) = \rho (\om) - \rho_{free} (\om)
\eeq
as an effective Casimir density of states, with normalization\footnote{As we already discussed,
there is a one to one correspondence between the modes of the constrained and the free systems. However,
we cannot use that correspondence to break (\ref{c}), which is supposedly a convergent integral, 
into the difference of separate integrals over $\rho$ and $\rho_{free}$, which are equal (due to the 
correspondence between states), but infinite.}
\beq\label{c}
\cac = \int\limits_0^\infty d\om\, \delta\rho (\om)\,.
\eeq
Following Section 6 of \cite{baliandup}, we define an average frequency $\bar\om$
through
\beq\label{aveomega}
\log\bar\om = {1\over\cac} \int\limits_0^\infty d\om\, \delta\rho (\om) \log\om\,.
\eeq
In terms of these quantities, we can write (\ref{classentropy1}) as 
\beq\label{classentropy}
\lim_{\frac{T}{T_c} >> 1} \cs_c = \cac\left(1-\log\left(\beta\hbar\bar\om\right)\right)\,.
\eeq
 
Balian and Duplantier were able to show that the normalization $\cac$
associated with electromagnetic fluctuations is 
\beq\label{c1}
\cac = {1\over 128\pi}\int_{\rm boundaries} d^2\si \left ({3\over R^2_1} + {3\over R^2_2} + {2\over R_1R_2}\right)
- n\,, 
\eeq
where $R_1$ and $R_2$ are the local curvature radii of the boundary, and $n$ is the genus of the boundary. 
Using the Gauss-Bonnet formula
\beq\label{gaussbonnet}
n = n_c - {1\over 4\pi}\int_{\rm boundaries} {d^2\si \over R_1R_2}\,,
\eeq
where $n_c$ is the number of connected pieces of the boundary, we can write (\ref{c1}) alternatively
as 
\beq\label{c2}
\cac = {1\over 128\pi}\int_{\rm boundaries} d^2\si \left ({3\over R^2_1} + {3\over R^2_2} + {34\over R_1R_2}\right)
- n_c\,.
\eeq

An obvious feature of (\ref{c1}) and (\ref{c2}) is the absence of a term 
proportional to the area 
$$\caa = \int_{\rm boundaries} d^2\si \,.$$
This is peculiar to electromagnetic fluctuations, due to 
cancellation between electric and magnetic modes.\footnote{For more details, 
see Section 4 (and in particular, Sections 4.2 and 4.3) in \cite{baliandup}.} 
In the case of fluctuations of a relativistic massless scalar field, for example, 
which we discuss in Section 3, Balian and Bloch\footnote{See Eq.(1.7) in the 
first reference cited in \cite{balian}. Our $\rho (\om)$ is $2\om$ times the
expression in \cite{balian}.} showed that 
\beq\label{deltarhoscalar}
\delta\rho (\om) = \rho (\om) - \rho_{free} (\om) = \mp {{\caa}\om\over 8\pi} 
+ {1\over 12\pi^2}\int_{\rm boundaries} d^2\si \left ({1\over R_1} + {1\over 
R_2}\right) + {\cO}\left({1\over\om}\right)\,,
\eeq
where the $`-'$ sign occurs for Dirichlet boundary conditions and the $`+'$ 
sign, for Neumann boundary conditions.\footnote{The fact that Dirichlet and 
Neumann boundary conditions induce opposite signs for the area 
term in (\ref{deltarhoscalar}) is essentially the reason why, in the case of 
electromagnetic fluctuations, the area term contributions of electric and 
magnetic modes cancel each other.} Thus, the normalization $\cac$ for 
scalar field fluctuations, will have a (divergent, due to the $\om$ integration) 
area contribution.

From these considerations we see that the Casimir entropy in the classical limit, as given by 
(\ref{classentropy}), depends on the geometry of the boundaries, but it also depends logarithmically on temperature. It is not just a geometric dependent constant as required by ``Kirchhoff's theorm''.
The Casimir entropy (\ref{classentropy}) can be derived in the usual manner from the Casimir free energy 
\beq\label{semiclassical}
\cf_c\simeq -\cac T\log \left({k_B T\over \hbar\bar\om}\right)\,,
\eeq
which is precisely the high temperature expression obtained in \cite{baliandup} via the mutiple scattering 
expansion. We can interpret (\ref{semiclassical}) as the high temperature limit of the free energy of 
$\cac$ non-interacting harmonic oscillators of frequency $\bar\om$. 
From (\ref{semiclassical}) we thus find that the Casimir energy is
\beq\label{semiclassicalenergy}
\ce_c = {\pa(\beta\cf_c)\over\pa\beta} = \cac T\,,
\eeq
and it does not vanish. It is of course a statement of classical equipartition: according to (\ref{c}) we can 
interpret $\cac$ as the number of additional modes of finite frequency created by introducing the conducting 
boundaries, each of which carries thermal energy $k_BT$. 

How do we reconcile these high temperature behaviors of $\cs_c, \ce_c$ and $\cf_c$ with our expectations of 
the classical limit? It is at this point that the relative thermodynamic quantities which we defined above
become useful: It is clear from the explicit formulas (\ref{c1}) and (\ref{c2}) that $\cac$ depends only on the 
geometrical features of each individual connected piece of the boundary, as well as on the overall topology of 
their arrangement in space, but {\em not} on the relative distances and orientations of these components.
(On the other hand, it is clear from \cite{baliandup} that the average frequency $\bar\om$, does depend on the
the relative distances and orientations of these components; otherwise, there will be no Casimir 
forces between them.) Thus, we immediately conclude that the relative 
Casimir entropy (\ref{relentropy}) tends in the classical limit to  
\beq\label{classrelentropy}
\lim_{\frac{T}{T_c} >> 1} \cs_c^{rel} = \cac\log\left({\bar\om (\infty)\over\bar\om}\right)
\eeq
(where $\bar\om(\infty)$ is the average frequency (\ref{aveomega}) of the system with all boundaries 
far apart),
which indeed depends only on the geometry of the boundaries. Similarly, the {\em relative} Casimir 
energy (\ref{relenergy}), $\ce_c^{rel} (T) = \ce_c (T) - \ce_c^\infty (T)$, vanishes in the classical limit, 
because the constrained system in which all the same disjoint boundaries are infinitely far apart, 
has the same coefficient $\cac$. ``Kirchhoff's theorem'' (\ref{kirchoff}) holds.

It follows then that the relative Casimir {\it free} energy $\cf_c^{rel}$, namely, the difference 
\beq\label{relfree}
\cf_c^{rel} = \cf_c - \cf_c^\infty
\eeq 
between the Casimir free energy of the constrained system and a system with the same boundaries infinitely 
far apart, is proportional to the temperature in the classical limit, 
\beq\label{relfree1}
\cf_c^{rel} = -T\cs^{rel}_c\,.
\eeq
Hence, the Casimir forces between disjoint boundaries 
are purely entropic in the classical limit, as was mentioned above.
This reasoning is born out by the explicit calculations presented in the next section and in 
Appendices A and C.

An interesting point to note is that sign of $\cac$ can be either positive or negative\footnote{The 
difference in mode densities $\delta \rho (\om)$ (\ref{casdensity}) need not be positive definite.}, as should be 
clear from the explicit expressions (\ref{c1}) and (\ref{c2})\cite{baliandup}. For example, a single toral 
boundary will have a positive $\cac$ if the hole in the center of the torus is very small (i.e., if the 
torus is a very fat tube), or if the torus is very narrow compared to the radius of the hole 
(i.e., the torus is a very thin tube). A torus inbetween these domains, will have a negative $\cac$. 
(For more details see \cite{baliandup}.) This means that the classical relative entropy $\cs_c^{rel}$ 
(\ref{classrelentropy}) (being a difference of entropies) need not be positive, and, also, that the Casimir forces
between the disjoint boundaries may be either attractive or repulsive, depending on the geometry (and topology)
of the boundary pieces. (The signs of $\cs_c^{rel}$ and of the forces depend of course also on $\bar\om$.)

We are now in a position to interpret the zero point energy (zpe) as a contribution 
necessary to assure that at high temperatures , the energy, $U$, is a 
function 
of $T$ only as, indeed,  was noted long ago (in 1913) by Einstein and Stern 
(\cite{milonni}, p.2): without it the energy will depend on the boundaries.  
Alternatively, if we assume the validity of 
``Kirchhoff's theorem'' at 
high temperatures, we may deduce the zero temperature (relative) Casimir energy as 
follows.

Let us write the energy per mode $k$ of the radiation field (constrained between several disjoint boundaries)
as 
\begin{equation}\label{utc}
E_k (T;T_c) = E_k (0;T_c) + E_k' (T;T_c)
\end{equation}
where $E_k'$ is the energy held in the mode without the zpe. 
For an allowed mode $k$
\begin{equation}
E_k' (T;T_c) = \frac{\hbar \omega_k}{\exp(\beta \hbar \omega_k) - 1}
\end{equation}
(see (\ref{aenergy})). Similarly, we denote the analogous quantities for the system with the same boundaries 
far apart as $E_k^\infty(T)$ and $E_k^{'\infty}(T)$, respectively. The relative Casimir energy is then
\begin{equation}
\ce_c^{rel} (T;T_c) = \sum_{k} E_k (T;T_c) - \sum_{k} E_k^\infty (T).
\end{equation}
Use of  (\ref{utc}) ( i.e., separating the thermal energy from the vacuum's) and 
``Kirchhoff's theorem'' (\ref{kirchoff}) implies
\begin{equation}
\lim_{T/{T_c} \rightarrow \infty} {\left[ \sum_{k} E_k'(T;T_c) - 
\sum_{k} E_k^{'\infty} (T) +  \ce_c^{rel}(0)\right]} = 0.
\end{equation}
with $\ce_c^{rel}(0)$ being the relative Casimir energy at zero temperature (which, of course,
for the parallel plates geometry, coincides with (\ref{casenphys})). Hence
\begin{equation}
\lim_{T/{T_c} \rightarrow \infty} \left[ \sum_{k} E_k'(T;T_c) -
 \sum_{k} E_k^{'\infty}(T) \right] = -  \ce_c^{rel}(0).
\end{equation}
Now both terms in the square bracket are readily calculable - their 
temperature dependence assures convergence - and thus yield the relative Casimir 
energy at $T=0$ without recourse to a (perhaps) objectionable regularization scheme.

To summarize, the validity, in the classical limit, of ``Kirchhoff's theorem'' (\ref{kirchoff}) implies 
that, in this limit, the relative Casimir entropy is independent of the temperature $T$ (and obviously, 
of Planck's constant $h$), and is thus a geometry dependent quantity that characterizes the constraints. 
Further, its validity allows a novel way to evaluate the $T=0$ value of the Casimir energy, as we 
discussed above.

\newpage

\section{Classical Limit for Arbitrary Geometry}

Our discussion of the classical limit of the Casimir effect concentrated so far on one concrete example, namely, the
parallel plates geometry. However, the intuitive physical argument presented in the previous section 
for the vanishing of the Casimir energy in the classical limit is quite general and does not seem to depend in any 
specific manner on the particular geometric details of the parallel plates system. Thus, we expect that it should be possible 
to supplement this argument by an explicit calculation of the classical limit of the Casimir effect for a cavity of 
arbitrary shape. In this section we carry out such a calculation, using path integration techniques. Some technical 
details are deferred to Appendices B-D. From this point onward, we reset $\hbar, c$ and $k_B$ to 1, unless 
explicitly stated otherwise.

\subsection{Zero temperature analysis}
In order to set up the formalism we start with the zero temperature case. For simplicity, we concentrate on a real massless
scalar field with action 
\begin{equation}\label{action}
S = \int d^4 x~ {1\over 2} ~\left(\partial_\mu\phi(x)\right)^2\,, 
\end{equation}
instead of the electromagnetic field. (Discussion of the electrodynamic Casimir effect follows essentially the same lines.)
We assume that the field is confined within a static cavity of an arbitrary shape and is constrained to vanish on its
boundary $\Sigma $, which may consist of several disjoint pieces. In addition, we assume that the curvature of $\Sigma$ is regular everywhere. 

\subsubsection{Sharp boundary conditions over all modes}
Following \cite{kardar, nussinov1, nussinov2}, we parametrize the world-volume swept out by our static $\Sigma $ as 
\begin{equation}\label{embed}
x^0=t;~~ {\bf x} = {\bf x}({\bf u})
\end{equation}
where ${\bf u} = (u_1,u_2)$ are internal coordinates on $\Sigma $.

The in-vacuum to out-vacuum transition amplitude ${\cal Z}$ associated with (\ref{action}) with $\phi(x)$ constrained to vanish 
on (\ref{embed}) is given by 
\begin{equation}\label{pathzc}
{\cal Z}_c = {1\over {\cal Z}_0} \int {\cal D}\phi \prod_{x\in\Sigma}~\delta \left[\phi(t, {\bf x}({\bf u}))\right ]~
{\rm exp} {i\over 2}\int d^4 x \,(\partial_\mu\phi)^2
\end{equation}
with normalization
\begin{equation}\label{z0}
{\cal Z}_0 = \int {\cal D}\phi~ {\rm exp} {i\over 2}\int d^4 x \,(\partial_\mu\phi)^2\,.
\end{equation}
We chose to normalize (\ref{pathzc}) relative to the unconstrained system, since, after all, we are interested
only in the difference of the ground state energies.

Following \cite{kardar} we write 
\begin{equation}\label{constraint}
\prod_{x\in\Sigma}~\delta \left[\phi(t, {\bf x}({\bf u}))\right ] = \int {\cal D}\psi (t,{\bf u})~ 
{\rm exp} ~i\int\, dt\, \int_\Sigma d^2 u \sqrt{g} \psi(t,{\bf u})\,\phi(t, {\bf x}({\bf u}))\,.
\end{equation}
Here 
\beq\label{detg}
g = \left({\partial{\bf x}\over \partial u_1}\right)^2 \left({\partial{\bf x}
\over \partial u_2}\right)^2 - \left({\partial{\bf x}\over \partial u_1}\cdot {\partial{\bf x}\over \partial u_2}\right)^2
\eeq
is the determinant of the induced metric on $\Sigma $.

Since $\Si$ is static, it is convenient to resolve $\phi$ and $\psi$ into frequency modes as
\begin{equation}\label{modes}
\phi(t,{\bf x}) = \int\limits_{-\infty}^{\infty} {d\om\over 2\pi}~\phi_\omega ({\bf x})~e^{-i\omega t}\quad {\rm and}\quad
\psi(t,{\bf x}) = \int\limits_{-\infty}^{\infty} {d\om\over 2\pi}~\psi_\omega ({\bf x})~e^{-i\omega t}
\end{equation}
(with $\phi_\om^* = \phi_{-\om}$, and similarly for $\psi$).

Substituting (\ref{constraint}) in (\ref{pathzc}) and using (\ref{modes}) we have 
\begin{equation}\label{pathzc1}
{\cal Z}_c = {1\over {\cal Z}_0} \int \prod_{\om\geq 0} {\cal D}\phi_\om {\cal D}\psi_\om~ 
{\rm exp} {i\over 2}\int\limits_{-\infty}^{\infty} {d\om\over2\pi}\left[\int d^3 x~\phi_\om^*~(\omega^2 + \nabla^2)\phi_\om +\int_{\Sigma} 
d^2 u ~\sqrt{g}~(\psi_\om^*\,\phi_\om +{\rm h.c.})\right]\,.
\end{equation}

We may summarize (\ref{constraint}) and (\ref{pathzc1}) by saying that the $\psi_\om$ act as sources for the $\phi^*_\om$, which 
live on $\Sigma$. Integration over these sources constrains all modes $\phi_\om$ (i.e., low frequency modes as well as high 
frequency modes) to vanish on $\Sigma$. This is, obviously, an idealized situation, e.g., if we try to take 
(\ref{action}) more 
seriously as a simplified version of the electrodynamic case. Indeed, it is well known that 
real metals are transparent to radiation of frequency $\omega\geq \omega_p$, where $\omega_p$ is the plasma frequency 
of the metal in question (which typically lies in the UV region). Low frequency modes (i.e., modes with $\om<\om_p$), on 
the other hand, penetrate the surface of the conductor only to within a distance of order $1/\om_p$ (the ``skin" depth) 
and are otherwise excluded from the bulk of the conductor.  Thus, as is well known, the ``perfect conductor'' boundary condition 
that all modes $\phi_\om ({\bf x})$ vanish on $\Si$, is an effective description (or a consequence) of the microscopic 
interactions of $\phi_\om$ and its sources $\psi_\om$ in the medium, where we assume $\om_p\rightarrow\infty$. The plasma frequency $\om_p$ serves as an effective UV 
cutoff frequency for the Casimir effect. In the next sub-section we will treat this issue in some detail. One of the benefits of 
doing so, besides gaining better physical understanding, is that the formalism we follow will enable us to calculate deviations from 
a perfect conductor behavior in a straightforward manner.

\subsubsection{More realistic interactions at the boundary}
Thus, to make our model more realistic, we should really think of the conducting surface $\Sigma$ as having some 
thickness $w$ of a few $1/\om_p$. (In doing so we make the tacit assumption that local curvature radii everywhere
on $\Si$ are much larger than $1/\om_p$.) We denote this thickened three dimensional conductor $\cb$. Only modes of 
frequency $\om<\om_p$ are constrained to decay into $\cb$. Modes of frequency $\omega > \omega_p$ will not be subjected 
to such a boundary condition. This physical behavior should be taken into account in (\ref{pathzc1}). 

Thus, we replace (\ref{action}) by the action 
\begin{equation}\label{action1}
S = {1\over 2}\int d^3 x\int {d\om\over 2\pi}~ \phi_\om^*({\bf x})\left[\om^2 \epsilon(\om, {\bf x}) + \nabla^2\right]
\phi_\om({\bf x})\,, 
\end{equation}
where, in analogy with electrodynamics, we introduced the dielectric function
\beq\label{dielectric}
\epsilon(\om, {\bf x}) = \left\{\begin{array}{cc} 1 + 4\pi\chi(\om), & {\bf x}\in\cb\\{} & {}\\ 1, & {\bf x}\notin\cb
\end{array}\right.  
\eeq
with susceptibility $\chi(\om)$, which encodes the polarization effects due to the 
$\phi_\om -\psi_\om$ interactions in $\cb$. 

As discussed in the previous section, the cavity bounded by $\cb$ has an IR cutoff frequency 
$\om_c$, which obviously sets the scale for frequencies that dominate the Casimir effect. We assume 
henceforth that the ratio 
$\om_p/\om_c$ is large but finite so as to leave a wide window open between the two cutoff frequencies.
For the parallel plate capacitor of Section 1, with plate separation $d$, we have of course 
\beq\label{omegac}
\om_c = \pi/d\,.
\eeq 
Then the condition $\om_p/\om_c = \om_p d/\pi >> 1$ means that we are allowed to use the 
perfect conductor boundary conditions only for plate separation $d$ in the 
range
\beq\label{drange}
d >> 0.06\mu{\rm m}\,\left( {10{\rm eV}\over \hbar\om_p}\right)\,. 
\eeq
If we decrease the plates' separation $d$ below what is allowed by (\ref{drange}), $\om_c$ would tend to $\om_p$
until it would eventually cross it over. Field modes around $\om_c$ would penetrate yet deeper into the
boundary, until it would become effectively transparent to these modes, rendering the boundary 
conditions less effective. Thus, we expect that if our modified model (as given 
by (\ref{action1}) and (\ref{dielectric})) could still be trusted when $\om_p/\om_c \leq 1$, 
then the $-1/d^4$ behavior of the Casimir force (\ref{casimirforce}) would crossover to a less
divergent behavior. In fact, assuming (\ref{chi}) for $\chi (\om)$, we can show that the Casimir force behaves 
like $-1/d^3$ for $d$ below the range allowed by (\ref{drange}) \cite{lifshitz} (see also \cite{milonni}, p.230).

Current experiments are performed (albeit not in the parallel plate geometry) 
with $d\geq 0.1\mu$m\cite{prl}. Obviously, it would be interesting to check experimentally 
(when technology would allow performing such experiments) how the Casimir force behaves at such 
much smaller distances (which are yet large enough on the atomic scale). 

In addition to (\ref{drange}), we assume that $\om_c$ is large enough, so that for 
$\om\geq\om_c$ we may neglect all resonances and dissipative effects encoded in $\chi$, and thus 
write\footnote{Following elementary dispersion theory, we may take (as the simplest example) 
$\chi(\om) = -\gam\si/\om (\om +i\gam)$. The simple pole at $\om=0$ corresponds, of course, 
to $\cb$ being a conductor, with conductivity $\si$ and damping coefficient $\gam = \om_p^2/4\pi\si$. 
Our second assumption is simply that $\om_c >> \gam$. More generally, we assume that $\om_c$ is much 
bigger than the highest resonance frequency and largest damping coefficient 
appearing in $\chi$. Finally, note that the analytic structure of $\chi (\om)$ insures that the path integral over (\ref{action1}) 
is well defined, namely, Re$(iS)<0$ always.}
\beq\label{chi}
4\pi\chi(\om) = -{\om_p^2\over \om^2}\,,
\eeq
in direct analogy with the theory of dielectrics.
(In reality, this corresponds to frequencies in the infrared and higher. In the case of the parallel plate
capacitor, tuning $\om_c$ to the infrared, means plate separation $d$ around $1\mu$m.)

With (\ref{chi}) for $\chi$, the action (\ref{action1}) is equivalent to 
\begin{equation}\label{action2}
S = {1\over 2}\int d^3 x\int {d\om\over 2\pi}~ \phi_\om^*({\bf x})\left[\om^2 + \nabla^2 - m^2({\bf x})\right]
\phi_\om({\bf x})
\end{equation}
with 
\beq\label{mass}
m^2({\bf x}) = \left\{\begin{array}{cc} \om_p^2, & {\bf x}\in\cb\\{} & {}\\ 0, & {\bf x}\notin\cb\,.
\end{array}\right.  
\eeq
$\phi$ develops a finite mass $\om_p$ inside $\cb$ and is thus expelled from it.\footnote{Long ago, Anderson \cite{anderson}
pointed out that the electrodynamical analogs of (\ref{action2}) and (\ref{mass}) may serve as a low energy, non-relativistic
example of dynamical mass generation of gauge fields, analogous to mass generation in the Schwinger model or to the Higgs phenomenon.} 
Moreover, we see from (\ref{action2}) that in the limit $\om_p\rightarrow\infty$ of a perfect conductor, $\phi_\om({\bf x})$ tends 
to vanish on $\Si$.

$\epsilon(\om)-1 = 4\pi\chi(\om)$ in (\ref{action1}) is, of course, the Green's function (at ${\bf k}=0$) of the source field
$\psi$ (already mentioned after (\ref{pathzc1})), that lives in $\cb$. Thus, (\ref{action1}) is equivalent to 
\begin{equation}\label{action3}
\sef = {1\over 2}\int {d\om\over 2\pi}~\left\{\int d^3 x~ \phi_\om^*\left[\om^2 + \nabla^2\right]
\phi_\om + \int_{\cb} d^3 x~\left[ {-\psi_\om^*\psi_\om\over 4\pi\om^2\chi(\om)}  + (\psi_\om^*\,\phi_\om +{\rm h.c.})\right]\right\}\,.
\end{equation}
Thus, we replace (\ref{pathzc1}) by 
\begin{equation}\label{pathzc2}
{\cal Z}_c = {1\over {\cal Z}_0 {\cal Z}_\psi} \int \prod_{\om\geq 0} {\cal D}\phi_\om {\cal D}\psi_\om~ 
{\rm exp} ~i\sef, 
\end{equation}
with
\beq\label{zpsi}
{\cal Z}_\psi = \int \prod_{\om\geq 0} {\cal D}\psi_\om~ {\rm exp} {i\over 2}\int\limits_{-\infty}^{\infty} 
{d\om\over 2\pi}\int_{\cb} d^3 x~{-\psi_\om^*\psi_\om\over 4\pi\om^2\chi(\om)}\,.
\eeq

With (\ref{chi}), the quadratic piece in $\psi$ in (\ref{action3}) is $+|\psi_\om|^2 /\om_p^2$. In the limit $\om_p\rightarrow\infty$
of a perfect conductor, $+|\psi_\om|^2 /\om_p^2$ goes away, the thickened conducting surface $\cb$ shrinks 
back into $\Si$, and (\ref{pathzc2}) reduces to (\ref{pathzc1}) (as we already commented following (\ref{mass})).

Observe that in our discussion of the Casimir effect in Sections 1 and 2, and in this section up to (\ref{pathzc1}), 
we did not make any reference to the fine structure constant $\alpha$ (or its analog, associated with the $\phi\psi$ 
coupling in $\cb$). However, since $\om_p^2$ is proportional to $\alpha$, it rightfully sneaks into our discussion 
following the modified action (\ref{action1}), or equivalently, (\ref{action3}). Given this, we should, in principle, include the $\phi\psi$ 
interaction (or analogously, the photon-electron interactions) also in the bulk part of the action in (\ref{action3}), taking into 
account the full dynamics of the source field $\psi$. This would imply using below the renormalized $\phi$ 
(or photon) propagator in the bulk (for such radiative corrections in the Casimir effect see, e.g.,\cite{bordag}). However, such 
effects are small and we ignore them in this paper.

Eqs. (\ref{action3}) and (\ref{pathzc2}) capture some aspects of the dynamics of $\psi$ 
(and thus, in a sense, of the dynamics of the boundaries where $\psi$ lives), rendering our formalism more 
realistic as it relaxes the constraints on higher frequency modes and thus regularizes the theory. One aspect of this
regularization is that the two dimensional surface $\Si$ becomes ``fuzzy" and swells into a shell $\cb$
of finite width. 
In this context we should also mention a recent work by Ford and Svaiter \cite{ford} which also takes boundary 
dynamics into consideration, but in a different (and presumably complementary) way. In \cite{ford} the boundary as a 
whole is quantized - its location serves as a collective heavy coordinate which sits in its ground state (at $T=0$) 
and is thus not sharply defined. This quantum uncertainty in the boundary location also regularizes the theory since it 
cures an old problem having to do with divergences of the energy density near sharply defined boundaries (see, e.g. the work by 
Candelas and Deutsch (1979) in \cite{dynamic}). The idea is simple: if the field vanishes on a sharply defined surface, 
its momentum at all points on the surface is unbounded due to the uncertainty principle, rendering the field energy density 
singular on the surface. If the location of the boundary has zero point fluctuations, the field at the (``fuzzy'') boundary is not 
sharply constrained, and the field energy density is no longer singular. 

In the spirit of \cite{kardar, nussinov1, nussinov2} we would like now to integrate $\phi$ out of (\ref{pathzc2}). To this end it is useful to 
extend the definition of the sources $\psi_\om$ from $\cb$ into all space as 
\begin{equation}\label{psisource}
\psi_\om({\bf x}) \rightarrow \left\{\begin{array}{cc} \psi_\om ({\bf x}), & {\bf x}\in\cb\\{} & {}\\ 0, & {\bf x}\notin\cb\,.
\end{array}\right.  
\end{equation}
In terms of these extended sources, clearly all spatial integrals in (\ref{action3}) involving $\phi$, including 
the $\phi\psi$ interaction term, are carried over all space. Thus, integrating $\phi$ out is straightforward and produces a factor 
${\cal Z}_0$, which cancels the corresponding normalization factor in (\ref{pathzc2}).
We end up with 
\begin{equation}\label{pathzc3}
{\cal Z}_c = {1\over {\cal Z}_\psi}\int \prod_{\om\geq 0} {\cal D}\psi_\om~ {\rm exp}~\left\{
{-i\over 2}\int {d\om\over 2\pi}\int\limits_{x, x'\in\cb} d^3 x d^3 x'~ \psi_\om^*({\bf x}) \left[ {\delta^{(3)}({\bf x - x'})
\over \om_p^2} + G_\om({\bf x}, {\bf x'})\right]\psi_\om({\bf x'})\right\}\,,
\end{equation}
where 
\begin{eqnarray}\label{gkernel}
G_\om({\bf x}_1, {\bf x}_2) = <{\bf x}_1|~{1\over {-\omega^2} - \nabla^2 +i\epsilon}|{\bf x}_2>
= {e^ {- i|\omega|~|{\bf x}_1 - {\bf x}_2|}\over 4\pi~|{\bf x}_1 - {\bf x}_2|}\,.
\end{eqnarray}

In order to proceed, we recall at this point that $\cb$ is essentially the thickened two dimensional conducting surface $\Si$, 
with normal width $w$ of a few skin depths $l/\om_p\,\,(l\sim {\cal O}(1))$. Thus, we parametrize $\cb$ by 
$$ {\bf x} = {\bf x}({\bf u}, w)\,,$$
where ${\bf u}$ are the two internal coordinates of $\Si$ (see ({\ref{embed})), and $w$ is the coordinate in the direction 
normal to $\Si$. We are interested, of course, in the limit of large $\om_p$, or very small width $w$. Clearly, in this limit $\psi_\om$ 
(as well as $\phi_\om$) do not fluctuate in the normal direction. Consequently we may write the $\psi$-quadratic action in 
(\ref{pathzc3}) as 
\beq\label{psiquad}
S_\psi = {1\over 2}\int {d\om\over 2\pi}\int\limits_{\Si} d^2 u~d^2 u'~ \psi_\om^*({\bf u}) 
\left[ {l\sqrt{g({\bf u})}\over\om_p^3}~\delta^{(2)}({\bf u - u'}) + \left({l\over \om_p}\right)^2 {\cal M}_\om({\bf u}, {\bf u'})\right]
\psi_\om({\bf u'})
\eeq
where 
\beqra\label{mkernel}
{\cal M}_\om({\bf u}, {\bf u'}) &=& \sqrt{g({\bf u})}~G_\om({\bf x(u)}, {\bf x(u')})~\sqrt{g ({\bf u'})}\nonumber\\{}\nonumber\\
&=& \sqrt{g({\bf u})}~{e^ {-i |\omega| ~|{\bf x}({\bf u}) - {\bf x}({\bf u'})|}\over 4\pi~
|{\bf x}({\bf u}) - {\bf x}({\bf u'})|}~\sqrt{g({\bf u'})}
\end{eqnarray}
is the kernel of a symmetric operator $\hat{\cal M}_\om$. 

We see that ${\cal Z}_c$ is expressed entirely in terms of the sources $\psi_\om$. We may now readily integrate over the 
$\psi_\om$ and obtain 
\begin{eqnarray}\label{zfinal}
{\cal Z}_c &=& \prod_{\om\geq 0}\,'~ {1\over {\rm det}\left[{\bf 1}_{\Si} + l\om_p\hat{\cal M}_\om\right]}
\nonumber\\{}\nonumber\\
&=& {\rm exp}~\left\{-{{\cal T}\over 2}\int\limits_{-\infty}^{\infty} {d\om\over 2\pi} \rmtr\log~\left[  
{\bf 1}_{\Si} + l\om_p\hat{\cal M}_\om\right]\right\}\,.
\end{eqnarray}
Here ${\cal T}$ (implicit in the product in the first line of (\ref{zfinal})), is a temporal IR cutoff which arises due to 
time translational invariance. The prime to the right of the product symbol after the first 
equality in (\ref{zfinal}) is to remind us that $\psi_{(\om=0)}$ is a real field, and thus integration over it yields a square root 
of the appropriate determinant. This is taken care of in the last expression in (\ref{zfinal}), where we used $\hat{\cal M}_{-\om} = 
\hat{\cal M}_{\om}.$ The Casimir energy is thus
\beq\label{casenergy}
{\cal E}_c = {i\over {\cal T}}\,\log {\cal Z}_c = -{i\over 2}\int\limits_{-\infty}^{\infty} {d\om\over 2\pi} \rmtr\log~\left[  
{\bf 1}_{\Si} + l\om_p\hat{\cal M}_\om\right]\,.
\eeq
Eq. (\ref{casenergy}) is one of our main results and we will use it to infer the classical (high temperature) 
limit of the Casimir effect in the next sub-section.

Note that if $\Sigma$ is the union of several disjoint two dimensional surfaces $\Sigma_\alpha$ ($\alpha = 1, 2, \cdots N$), 
where $\Sigma_\alpha$ has local coordinates ${\bf u}^{(\alpha)}$ and induced metric $g^{(\alpha)}$, then obviously $\psi_\om$ is the sum
\begin{equation}\label{psisourcemany}
\psi_\om({\bf x}) = \sum_\alpha \psi_\om^{(\alpha)}({\bf x})
\end{equation}
of contributions from the individual $\Sigma_\alpha$'s (in obvious notation).
It is clear that in such a case, 
$\hat{\cal M}_\om$ is actually an $N\times N$ operator matrix $\hat{\cal M}_\om^{\alpha\beta}$ with kernel 
\begin{equation}\label{mkernel1}
{\cal M}_\om^{\alpha\beta}({\bf u}^{(\alpha)}, {\bf u}^{(\beta)}) = \sqrt{
g^{(\alpha)}({\bf u}^{(\alpha)})}~
<{\bf x}({\bf u}^{(\alpha)})|
~{1\over {-\omega^2} - \nabla^2 +i\epsilon}|{\bf x}({\bf u}^{(\beta)})>~\sqrt{g^{(\beta)}({\bf u}^{(\beta)})}\,.
\end{equation}
(\cite{kardar} actually discusses the most generic case where the field lives in $D+1$ dimensions and
is constrained to vanish on several manifolds $\Sigma_\alpha$ of various dimensions, some of which may be 
time dependent.) In particular, for the parallel plate geometry discussed in the previous sections ($\Sigma_1$ 
is the $z=0$ plane and $\Sigma_2$ is the $z=d$ plane), a straightforward calculation yields \cite{kardar,nussinov1, nussinov2}
\begin{equation}\label{mplates}
\hat{\cal M}_\om = \left(\begin{array}{cc} 1 & e^{-d\hat h_\om}\\ e^{-d\hat h_\om} & 1\end{array}\right) {1\over 2 \hat h_\om}
\end{equation}
with 
\begin{equation}\label{hhat}
\hat h_\om = \sqrt{-\omega^2 - \partial_x^2 - \partial_y^2 +i\epsilon}\,.
\end{equation}

If we let $\om_p$ tend to infinity in (\ref{casenergy}) we obtain the Casimir energy of the perfect 
conductor $\Si$, identical (up to an infinite additive constant proportional to the area of $\Si$) to the Casimir 
energy one may calculate directly from (\ref{pathzc1}). In particular, substituting 
(\ref{mplates}) in (\ref{casenergy}) yields (after proper regularization, and up to an obvious factor $1/2$) 
the Casimir energy (\ref{casen}).

However, keeping $\om_p$ large but finite, we may calculate corrections to the Casimir effect due to 
imperfect conductor behavior by expanding (\ref{casenergy}) in inverse powers of $\om_p$. Such calculations were made in \cite{mrs} 
using Green's function techniques. In Appendix C, following \cite{kardar, nussinov1, nussinov2} (and the formalism presented in this section), 
we briefly sketch the calculation of the Casimir force at zero and finite temperatures for the parallel plate capacitor using 
(\ref{mplates}). We also calculate in Appendix C the leading (in $1/\om_p$) correction to the Casimir effect due to imperfect conductor 
behavior for the parallel plate capacitor, which coincides with the results of \cite{mrs} and \cite{hargreaves}. 

\subsection{Finite temperature behavior and the Classical Limit}

The finite temperature case is obtained, in the usual manner, from the zero temperature case of the previous 
sub-section, by continuing $\om$ to imaginary values, where it can take only the discrete Matsubara 
frequencies $\om_n = 2\pi n /\beta$. Consequently, we replace continuous integrals 
$\int\limits_{-\infty}^{\infty} {d\om/ 2\pi} $ by discrete sums 
$(1/\beta)\sum_{n=-\infty}^{\infty}$ (with $\beta=1/T$). 

Obviously, since $\om_p$ plays the role of a UV cutoff, we must sum only over $|\omega_n| < \om_p$. 
We show in Appendix B that the physics of the electron Fermi gas in the metal actually warrants $\om_1 << \om_p$,
so that these discrete sums include many terms, and also that we may ignore the $T$ dependence of $\om_p$. 
These facts indicate that our cutoff procedure, using (\ref{dielectric}) and (\ref{chi}), is physically 
consistent also at finite and high temperatures.

Under the analytic continuation $\om\rightarrow -i\om$, the oscillatory 
exponentials in (\ref{gkernel}) and (\ref{mkernel}) are replaced by decaying exponentials. Thus for finite temperature
we replace ${\cal M}_\om$ by 
\beqra\label{mkernelt}
{\cal M}_n ({\bf u}, {\bf u'}) = \sqrt{g({\bf u})}~{e^ {-|\omega_n| ~|{\bf x}({\bf u}) - {\bf x}({\bf u'})|}\over 4\pi~
|{\bf x}({\bf u}) - {\bf x}({\bf u'})|}~\sqrt{g ({\bf u'})}\,,
\end{eqnarray}
(as well as a similar replacement for ${\cal M}_\om^{\alpha\beta}$ in (\ref{mkernel1})). We also introduce the 
obvious notational replacement $\hat{\cal M}_\om\rightarrow\hat{\cal M}_n$. Note that in the Euclidean domain, 
$\hat{\cal M}_n$ is not only symmetric, but also real.

In particular, note that at finite temperature we must replace (\ref{mplates}) and (\ref{hhat}) of the parallel plate 
capacitor by 
\begin{equation}\label{mplatest}
\hat{\cal M}_n = \left(\begin{array}{cc} 1 & e^{-d \hat h_n }\\ 
e^{-d \hat h_n } & 1\end{array}\right) {1\over 2 \hat h_n }\,,
\end{equation}
with $\hat h_n = \sqrt{\omega_n^2 - \partial_x^2 - \partial_y^2}$. 
  
Finally, we analytically continue (\ref{zfinal}) to imaginary $\om$ (and also replace ${\cal T}\rightarrow\beta$ in (\ref{zfinal}))
and obtain the partition function as
\begin{eqnarray}\label{zfinalt}
{\cal Z}_c &=& \prod_{n\geq 0}\,'~ {1\over {\rm det}\left[{\bf 1}_{\Si} + l\om_p\hat{\cal M}_n\right]}
\nonumber\\{}\nonumber\\
&=& {\rm exp}~\left\{-{1\over 2}\sum_{n=-\infty}^{\infty} \rmtr\log~\left[  
{\bf 1}_{\Si} + l\om_p\hat{\cal M}_n\right]\right\}\,.
\end{eqnarray}
Thus, the Casimir free energy (the analytic continuation of (\ref{casenergy})) is 
\beq\label{free}
{\cal F}_c = -{1\over\beta}\log\,{\cal Z}_c = {1\over 2\beta} \sum_{n=-\infty}^{\infty} 
\rmtr\log~\left[ {\bf 1}_{\Si} + l\om_p\hat{\cal M}_n\right]\,.
\eeq

\subsubsection {The Classical Limit }
Our main interest in this paper is of course the high temperature, classical limit 
of the Casimir effect. Recall from Section 1 (see in particular Fig. 1) that for the parallel plate capacitor
(and electromagnetic fields), the classical limit is 
$$\ce_c \rightarrow 0, \; \; \; \cf_c \rightarrow -T\cs_c, \; \; \; \cs_c \rightarrow 
\frac{\zeta(3)}{2^3 \pi}(\frac{L^2}{d^2}) $$  
(namely, (\ref{classicallimit}), which we repeat here for convenience). As discussed in Section 2, it is supposed 
to become valid for temperatures 
$T>>T_c$, where $T_c = \hbar c \pi/ k_B d$ (see (\ref{tc})) is a temperature scale defined by geometry.
Since $\om_p$ is our UV cutoff, it is clear that the relevant temperature range must be far below 
$\hbar\om_p$ (which corresponds to $\sim 10^5\,{}^o K$), 
\beq\label{temperaturerange}
k_BT_c < k_B T << \hbar\om_p\,,
\eeq
which is consistent with the condition $\om_p/\om_c >>1$ (Eq. (\ref{drange})) and also in accordance 
with the discussion in Appendix B.

According to the general discussion in Section 2, we would expect this behavior to be generic, and 
occur for cavities of arbitrary shape.  Thus, we expect that there should be a temperature scale, which we 
also denote by $T_c$, which is determined by the 
geometry of the cavity, in analogy with (\ref{tc}), and marks the onset of the classical limit.

In the rest of this section we prove this expectaion for cavities whose boundaries consist of several 
disjoint pieces, and also set a bound on $T_c$. Cavities bounded by a single connected boundary 
are in principle more difficult to handle in this respect, for reasons that we shall explain below.
In fact, for such cavities, we found logarithmic corrections to (\ref{classicallimit}), which seem to be 
consistent with the results of Balian and Duplantier \cite{baliandup}.

We see from (\ref{mkernelt}) that ${\cal M}_n({\bf u}_1, {\bf u}_2)$ decays exponentially fast, unless 
\beq\label{expfast}
\omega_n~|{\bf x}({\bf u}_1) - {\bf x}({\bf u}_2)|\leq 1\,.
\eeq 
In particular, it is clear from (\ref{mkernelt}) and (the analytic continuation of) (\ref{mkernel1}), that for $n\geq 1$
\beq\label{bounds}
{\cal M}_n^{\alpha\beta}({\bf u}^{(\alpha)}, {\bf u}^{(\beta)}) \leq
\sqrt{g^{(\alpha)}g^{(\beta)}}~
{e^{-\om_n d_{\alpha\beta}}\over 4\pi d_{\alpha\beta}}\,,\quad \alpha\neq\beta\,, 
\eeq
where $d_{\alpha\beta}$ is the minimal distance between the disjoint pieces 
$\Si_\alpha$ and $\Si_\beta$. 

Thus, if $T=\om_1/2\pi$ is such that $\om_1 d_{min} > 1$, where $d_{min}$ is the minimal $d_{\alpha\beta}$, 
then all matrix elements of ${\cal M}_n^{\alpha\beta}$ (for any pair $\alpha\neq\beta$) will be exponentially 
small. Below we show that these ${\cal M}_n^{\alpha\beta}$ make exponentially small contributions to the 
Casimir free energy already when 
\beq\label{bound1}
2\om_1 d_{min} > 1\,.
\eeq

The off-diagonal blocks of (\ref{mplatest}) are a clear example of 
this behavior, for which we can write the {\em operatorial} inequality
\beq\label{operatorialineq}
{e^{-d\hat h_n} \over 2 \hat h_n} \leq {e^{-\om_n d} \over 2 \om_n}{\bf 1}_\Si\,. 
\eeq
Indeed, as we show at the end of Appendix C, the high temperature behavior of 
(\ref{operatorialineq}) offers a very simple explanation of the exponential 
decay of the Casimir energy $\varepsilon_c(t)$ mentioned following (\ref{expfall}) 
(which we derive in Appendix A using a laborious mode summation), and of its corollary that the 
bulk number of modes per unit volume is independent of $d$.

The diagonal blocks of (\ref{mplatest}) obviously decay only as powers of $1/\om_n$. This is in 
fact true in general, as we now show. According to (\ref{expfast}) we should consider only neighboring 
points ${\bf x}({\bf u}) \simeq {\bf x}({\bf u}+{\bf a})$, which means that they must belong to the same
$\Sigma_\alpha$ (assuming finite $d_{min}$, of course). By continuity then $|{\bf x}({\bf u}) - {\bf x}({\bf u} + {\bf a})|^2 
\simeq \sum_{n,n'} g_{nn'} a_n a_{n'}$, where 
$g_{nn'} = {\partial {\bf x}\over \partial u_n}\cdot {\partial {\bf x}\over \partial u_n'}$ is the 
induced metric on $\Sigma$ mentioned above. From these considerations and from the 
representation\footnote{Note that (\ref{deltarep}) is the first term in the expansion 
${1\over 2\epsilon} e^{-|x|/\epsilon} = 
\left(\sum_{n=0}^\infty \epsilon^{2n}\pax^{2n}\right)\delta (x)$, which may be used to 
calculate higher order corrections to (\ref{mkernelimit}). Thus, the leading correction to 
(\ref{mkernelimit1}) is $\sim \nabla_{\bf u}^2/\om_n^3$.}
\begin{equation}\label{deltarep}
\delta (x) = \lim_{\epsilon\rightarrow 0^+} {1\over 2\epsilon} e^{-|x|/\epsilon}\,,
\end{equation}
we immediately conclude that 
\begin{equation}\label{mkernelimit}
{\cal M}_n({\bf u}_1, {\bf u}_2) 
{_{\displaystyle \longrightarrow}\atop
{\scriptstyle \omega_n\rightarrow\infty}
}\,\, {g({\bf u}_1)\over \pi\omega_n}~
\delta (|{\bf x}({\bf u}_1) - {\bf x}({\bf u}_2)|^2 ) = 
{\sqrt{g({\bf u}_1)}\over 2\pi\omega_n}~{\delta (|{\bf u}_1 - {\bf u}_2|)\over |{\bf u}_1 - {\bf u}_2|}\,.
\end{equation}

Thus, from the elementary formula
$$\int d^2u_1 d^2 u_2 {\sqrt{g({\bf u}_1)}\over 2\pi}~{\delta (|{\bf u}_1 - {\bf u}_2|)\over |{\bf u}_1 - {\bf u}_2|} f({\bf u_2}) = 
{1\over 2}\int d^2u_1 \sqrt{g({\bf u}_1)}~f({\bf u_1})$$
we conclude that   
\begin{equation}\label{mkernelimit1}
\hat{\cal M}_n 
{_{\displaystyle \longrightarrow}\atop
{\scriptstyle \omega_n\rightarrow\infty}
}\,\, {{\bf 1}_\Si\over 2\omega_n}\,.
\end{equation}

Eq. (\ref{mkernelimit1}) is an asymptotic formula for $\om_n\rightarrow\infty$, so for 
$1/d_{min}< \om_n << \om_p$ we might still have to include many subleading corrections to 
(\ref{mkernelimit1}), but the point is that the diagonal blocks decay only as powers of $1/\om_n$. 

To summarize, when $T$ is high enough so that (\ref{bound1}) holds, the off-diagonal blocks 
${\cal M}_n^{\alpha\beta}$ ($\alpha\neq\beta$) become exponentially small (for $n\neq 0$) , and are thus 
dominated by the power-like decaying blocks. Thus, it makes sense to split $\hat{\cal M}_n $~($n\neq 0$)
into its diagonal and off-diagonal pieces
\beq\label{split}
\hat{\cal M}_n = \hat{\cal M}_{n,d} +  \hat{\cal M}_{n,od} 
\eeq
(in obvious notations).

This splitting is not only according to the different asymptotic behaviors of 
$\hat\cm_{n,d}$ and $\hat\cm_{n,od}$ at large $\om_n$, but also according to their different physical 
interpretations:
Consider, for simplicity, a boundary $\Si$ which is the union of just two disjoint connected pieces $\Si_1$ and 
$\Si_2$. (The following discussion may be generalized immediately to an arbitrary set of disjoint boundaries 
$\Si_\alpha$.) Assume that $\Si_1$ and $\Si_2$ are infinitely separated at 
first, and thus do not interact. Clearly, in this case, the only non vanishing blocks of $\hat\cm$ are the 
diagonal ones $\hat\cm^{11}$ and $\hat\cm^{22}$, that encode the interaction of the field $\phi$ and 
the corresponding boundary $\Si_1$ and  $\Si_2$, respectively. By definition, $\hat\cm^{11}$  depends only on the 
details of $\Si_1$ and $\hat\cm^{22}$ depends only on the details of $\Si_2$. In addition, note that 
the matrix elements defining $\hat\cm^{11}$ and $\hat\cm^{22}$ (see(\ref{mkernel})) are manifestly invariant under 
three dimensional rotations and translations. Thus, the spectra of $\hat\cm^{11}$ and $\hat\cm^{22}$ are invariant 
under rigid rotations and translations of $\Si_1$ and $\Si_2$, respectively. Then, as $\Si_1$ and $\Si_2$ are 
brought closer together, they start interacting, and the off-diagonal blocks $\hat\cm^{12}$ and $\hat\cm^{21}$, 
which encode this mutual interaction, are no longer vanishing. As $\Si_1$ and $\Si_2$ are drawn closer, the off-diagonal blocks clearly become larger, while the diagonal blocks remain unchanged. (We could have invoked, of 
course, this argument to split $\hat\cm$ into diagonal and off-diagonal blocks also at $T=0$.)
The parallel plate capacitor, with $\hat\cm$ given by (\ref{mplatest}), provides a very simple example for this 
discussion. 

Given our system with boundaries $\Si_1$ and $\Si_2$, we may study its behavior under infinitesimal perturbations. 
We can perturb the system in two independent ways. One possibility is to slightly deform any one of the two 
boundaries (or both), but otherwise keep the state of the plates relative to each other unchanged. (An infinitesimal
self-similar rescaling is an example of such a deformation.) The second  possibility is to move the boundaries rigidly
relative to each other, that is, to make a relative translation or rotation, without changing their shapes.
According to the discussion above, perturbations belonging to the first possibility will induce changes 
in the diagonal blocks and also in the off-diagonal blocks. Perturbations belonging to the second possibility affect 
{\em only} the off-diagonal blocks. The gradients of the
free energy (\ref{free}) (or of the Casimir energy (\ref{casenergy}) at $T=0$) with respect to these deformations
yield the corresponding Casimir forces. Again, this is all very clear for the example of the two plate capacitor.

Computation of the energy associated exclusively with the diagonal blocks of $\hat\cm$, that is, the energy of the 
constrained system with $\Si_1$ and $\Si_2$ infinitely far apart, poses some conceptual difficulties.\footnote{We adapt here a discussion given in \cite{nussinov2}.} Consider $\Si_1$ for 
example. Its interaction with the field $\phi$ gives rise to the divergent (cutoff dependent) quantity 
$\rmtr\log\hat\cm^{11}$, which we may interpret as the energy needed to assemble $\Si_1$. This energy includes
the divergent self-energy of the ``charges'' that make $\Si_1$ (which is, by definition, a short distance effect), and interactions among patches of charges on much larger distance scales. The latter contain the Casimir
energy we sought. Then a conceptual question arises how to define a finite physically sensible Casimir energy,
separate it from the self-energy and thus extract it from the divergent $\rmtr\log\hat\cm^{11}$ of $\Si_1$. 
As an example, consider  $\Si_1$ to be a spherical shell of radius $R$ (an example that has 
been and is being studied extensively in the literature 
\cite{recently, lambib, boyer, mrssphere, miltonsphere, baliandup, brevikold, candelasphere, boyer2}). One may 
attempt to define its Casimir energy by comparing the shell 
with radius $R$ with another shell with a much larger radius. But as we inflate the shell of radius $R$ to 
compare it with the other shell, each point on the shell recedes from {\em all} the other points on the 
shell. In particular, infinitesimally separated points on the sphere will recede from one another too. Thus, 
inflating the sphere affects interactions over all distance scales, and in particular, on the scale of the UV cutoff - that is, the self-energy. On top of it all, and related to the difficulty of separating the Casimir energy 
from the self-energy, there is the more practical question of how to separate the Casimir force acting on our 
isolated $\Si_1$ from the total surface tension (and other similar effects), in order to measure it.
This is to be contrasted with what happens when we rigidly move $\Si_1$ and $\Si_2$ 
relative to each other. Then all points in $\Si_1$ move relative to all points on $\Si_2$, but there is no relative movement of points within the same $\Si_\alpha$, i.e., there is no change of the self-energy of each of the boundaries.
Such moves thus induce clear-cut changes in the Casimir energy of the system \cite{nussinov2}.

Having split $\hat\cm$ into diagonal and off-diagonal blocks according to (\ref{split}) we expand the free 
energy in powers of the off-diagonal part
\beqra\label{expand}
&&\rmtr\log~\left[ {\bf 1}_{\Si} + l\om_p\hat{\cal M}_n\right] = 
\rmtr\log~\left[ {\bf 1}_{\Si} + l\om_p\hat{\cal M}_{n,d}\right]\nonumber\\{}\nonumber\\
&&- {1\over 2} \rmtr\,\left[{l\om_p\over
{\bf 1}_{\Si} + l\om_p\hat{\cal M}_{n,d}}\,\hat{\cal M}_{n,od}\,{l\om_p\over
{\bf 1}_{\Si} + l\om_p\hat{\cal M}_{n,d}}\,\hat{\cal M}_{n,od}\right] + \cdots\,.
\eeqra
Note that the linear term in $\hat{\cal M}_{n,od}$ is absent: 
$$\rmtr\,\left[{l\om_p\over{\bf 1}_{\Si} + l\om_p\hat{\cal M}_{n,d}}\,\hat{\cal M}_{n,od}\right] = 0\,.$$
This expansion (\ref{expand}) is well behaved due to the exponential bound (\ref{bounds}) on 
$\hat{\cal M}_{n,od} $, as we demonstrate in Appendix D. 

Since the leading correction due to $\hat\cm_{n,od}$ is quadratic, we see from (\ref{bounds}) (and from Appendix D)
that off-diagonal contributions become exponentially small at temperatures $T$ such that $2\om_1 d_{min} > 1$.
This is (\ref{bound1}) alluded to above. Thus, we expect from (\ref{bound1}) 
that $T_c$ be bounded from below 
\beq\label{bound2}
T_c>{1\over 4\pi d_{min}}\,. 
\eeq
For the parallel plate capacitor, for example, with $T_c$ given by (\ref{tc}), we have $1/4\pi d = T_c/4\pi^2$.

Given a configuration of disjoint boundaries $\Si_\alpha$, an interesting quantity is the cost in free energy
(or cost in energy, at $T=0$) needed to assemble that configuration from an initial configuration of the same 
boundaries, but infinitely far apart. In other words, we would like to study the {\em relative} Casimir free  
energy (\ref{relfree}), which according to the discussion above is given by 
\beq\label{cost}
\cf_c^{rel} = {1\over 2\beta}\sum_{n=-\infty}^\infty\,\left\{\rmtr\log~\left[ {\bf 1}_{\Si} + 
l\om_p\hat{\cal M}_n\right]  - 
\rmtr\log~\left[ {\bf 1}_{\Si} + l\om_p\hat{\cal M}_{n,d}\right]\right\}\,,
\eeq
and is a well defined quantity, not plagued by infinite self-energy
terms that cancel between the two terms on the right hand side of (\ref{cost}). 

For the parallel plate capacitor, $\cf_c^{rel}$ (which we calculate in Appendix C), is, as expected, $1/2$ the 
Casimir free energy (\ref{freecphys}) which was derived in Section 1 for electromagnetic fluctuations. Thus, more 
generally, we may interpret (\ref{cost}) as the Casimir free energy (or the Casimir energy, at $T=0$) due to field modes that depend explicitly on the mutual arrangements of
the disjoint boundaries. (Such modes are to be contrasted, for example,  with modes that live inside any of the 
boundaries $\Si_\alpha$  which is a closed boundary enclosing a cavity. When this $\Si_\alpha$ is a perfect conductor,
the latter modes are disconnected from the outer world, and will thus contribute exclusively to 
$\hat\cm^{\alpha\alpha}$.)

It is $\cf_c^{rel}$ in (\ref{cost}) that in the classical limit $T>>T_c$ becomes proportional to $T$, with a 
proportionality coefficient (minus the relative Casimir entropy $\cs_c^{rel}$ (\ref{classrelentropy}) 
in the classical limit) which is determined by geometry, in a similar 
manner to (\ref{classicallimit}). 
Indeed, from (\ref{bounds}), (\ref{bound2}) and from the explicit bounds
(\ref{bound2ndorder}) and (\ref{expestimate}) in Appendix D, we see immediately that for $T > 1/(4\pi d_{min})$ all 
terms in (\ref{cost}) with $n\neq 0$ are exponentially small relative to the contribution of the zero mode. 
Thus, in this limit 
\beq\label{costzeromode}
\cf_c^{rel} \rightarrow {T\over 2} \left\{\rmtr\log~\left[ {\bf 1}_{\Si} + 
l\om_p\hat{\cal M}_0\right]  - 
\rmtr\log~\left[ {\bf 1}_{\Si} + l\om_p\hat{\cal M}_{0,d}\right]\right\}\,,
\eeq
with exponentially small corrections. This is the high temperature limit of the relative Casimir free energy, which becomes proportional to $T$. 
Thus, the relative Casimir energy  $\ce_c^{rel} (T)$ (\ref{relenergy})
(i.e., the Casimir energy due to field modes that depend explicitly on the mutual 
arrangements of the disjoint boundaries) tends to zero exponentially fast (at least as fast as 
$e^{ -2\om_1 d_{min}}$), and ``Kirchhoff's theorem'' (\ref{kirchoff}) holds. Thus,  
\beq\label{classicalentropy}
\cs_c^{rel} = - {1\over 2} \left\{\rmtr\log~\left[ {\bf 1}_{\Si} + 
l\om_p\hat{\cal M}_0\right]  - 
\rmtr\log~\left[ {\bf 1}_{\Si} + l\om_p\hat{\cal M}_{0,d}\right]\right\}
\eeq
is the asymptotic relative Casimir entropy (\ref{relentropy}) (i.e., the high temperature limit of the Casimir entropy due to field modes that depend explicitly on the mutual arrangements of
the disjoint boundaries). Expanding (\ref{classicalentropy}) formally in inverse powers of $l\om_p$ we obtain 
\beq\label{classicalentropy1}
\cs_c^{rel} = - {1\over 2} \rmtr\log~\left(\hat{\cal M}_0\hat{\cal M}_{0,d}^{-1}\right) - 
{1\over 2l\om_p}\rmtr~\left(\hat{\cal M}_0^{-1} - \hat{\cal M}_{0,d}^{-1}\right) + \cdots 
\eeq

It will be very interesting to prove the consistency of (\ref{classicalentropy1})
(at $\om_p\rightarrow\infty$, of course), and the analog of (\ref{classrelentropy}) for scalar field 
fluctuations. 

In Appendix C we calculate (\ref{costzeromode})-(\ref{classicalentropy1}) for fluctuations of the scalar 
field in a parallel plate geometry. As expected, we reproduce there half the classical limit values of the 
Casimir thermodynamic quantities given in (\ref{classicallimit}) for fluctuations of the electromagnetic 
field, and also write down explicitly the leading correction in $1/l\om_p$ to these quantities. 

We end this subsection with a remark concerning the high temperature behavior of the diagonal blocks of $\hat\cm$.
(We remind the reader that these diagonal blocks of $\hat\cm_n$ encode the interactions associated with
each surface separately, i.e., when the different surfaces are completely decoupled from each other.) 
For $\om_n$ large, $\hat\cm_n$ is dominated by its block diagonal part $\hat{\cal M}_{n,d}$ and approaches the 
asymptotic form (\ref{mkernelimit1}) ${\bf 1}_\Si/ 2\omega_n$ (up to power like corrections in $1/\om_n$), and is thus
a short distance effect on the surface, which must be related to the high temperature limit of the ``electromagnetic''
self-energy of the surface.  Thus, 
$\rmtr\log~\left[ {\bf 1}_{\Si} + l\om_p\hat{\cal M}_n\right]$ tends to 
\beq\label{asymptdiag}
\rmtr\log~\left[ \left( 1 + {l\om_p\over 2\om_n}\right){\bf 1}_{\Si} \right] = 
\caa\log~\left( 1 + {l\om_p\over 2\om_n}\right)\sim \caa\log~\left( {l\om_p\over 2\om_n}\right)\,,
\eeq
where $\caa$ is the total area of all surfaces. Thus, $\cf_c$, the unsubtracted free energy (\ref{free}), unlike
$\cf_c^{rel}$, will 
contain (roughly $\om_p/\om_c$) terms proportional to 
\beq\label{logterm}
\caa T\log\left( {l\om_p \over T}\right)\,.
\eeq 
This logarithmic behavior of $\cf_c$ (for the scalar field) is surely akin to the 
(renormalized, electromagnetic) high temperature Casimir free energy (\ref{semiclassical}). Since $\om_p$ is the 
UV cutoff scale, we expect to have renormalization counter terms such as 
$$T\log\left({\bar\om\over\om_p}\right)$$ that will renormalize (\ref{logterm}) into (\ref{semiclassical}).
It will be interesting to see this happen in a concrete calculation.

\newpage

\section{Concluding Remarks}

This paper gives the classical - the high temperature - limit of the 
Casimir effect. As a start we studied the simple example of the radiation field confined 
between two infinite parallel conducting plates separated by a distance d,
and found that the Casimir energy vanishes (up to exponentially small corrections)
in the classical limit.

From that observation we abstracted an argument that in the classical limit 
(defined to be temperatures such that the Rayleigh-Jeans, i.e., energy equipartition, 
theorem holds) ``Kirchhoff's theorem'' is valid - i.e., the energy density of the radiation 
field is a function of temperature only. This implies that, in this limit, the
Casimir energy vanishes. We showed that the zero point energy is required to 
assure the validity of ``Kirchhoff's theorem'' in the classical limit. 
Alternatively, assuming the validity of the theorem allows the evaluation of 
the Casimir energy 
at $T = 0$ without recourse to any regularization scheme. 
We noted, following \cite{milonni}, that these results were 
anticipated by Einstein and Stern in 1913, prior to the formulation of 
modern quantum mechanics and quantum field theory.
These authors noted that the high temperature expansion of what 
we now term Bose distribution function is ($\beta \hbar \omega << 1$)
$$ \frac{\hbar \omega}{\exp(\beta \hbar \omega) - 1} \rightarrow k_BT - 
\frac{1}{2}\hbar \omega + O(1/T).$$
Thus we have here a temperature independent term which contributes to 
the total energy. Its cancellation, i.e., the validity of 
``Kirchhoff's theorem'' in our presentation, requires a positive zero 
point energy, $+ \frac{1}{2}\hbar \omega$. Thus the removal of the zero 
point energy by considering a ``normally'' ordered Hamiltonian does not 
eliminate the need for zero point energy. The zero point energy is 
seen, according to  Einstein and Stern, to be required
for the correct classical limit or, in our terminology, for the validity
of ``Kirchhoff's theorem''. This stems here from the form of the Bose
distribution rather than the Hamiltonian.\\

The study of the (Casimir) free energy in the classical limit led to 
demonstrating that the Casimir entropy, is (in this limit) a temperature independent constant 
and reflects finite volume corrections in statistical physics. Thereby
the proportionality of the free energy to the temperature $T$, that is well known \cite{prl} is seen to be related to the Rayleigh 
Jeans limit \cite{boyer} rather indirectly - it is a consequence of the 
interesting result  that the Casimir force in the classical 
limit is purely entropic - in fact geometric. It is thence predicted that 
a large class of Casimir free energies, in the high T limit, will turn out to 
be proportional to the temperature - reflecting the purely entropic origin of 
the Casimir force in this limit. The crucial point is the existence of one
to one correspondence between states of the constrained and unconstrained 
systems.

We verified the expected classical limit for the Casimir effect due to a massless scalar 
field constrained by a set of disjoint boundaries of arbitrary geometries. This is achieved by subtracting 
off the contributions of the field correlation function that connect a boundary piece with itself without 
interaction with any other distinct boundary piece. In other words, our normalization is the partition functions with the 
boundary pieces infinitly far apart. This removes contributions that are independent of the inter boundary
pieces distances. Our analysis allowed us also to include corrections to the Casimir effect due to deviations 
from perfect conductor behavior.

\vskip0.5cm

{\bf Acknowledgements:}

\noindent This work was supported by the Technion VPR Fund, by the Fund for 
Promotion of Research at the Technion, by the Fund for Promotion of Sponsored 
Research at the Technion,  and by the Technion- Haifa University Joint 
Research Fund. JF's research has been supported in part by the Israeli 
Science Foundation grant number 307/98 (090-903). Special thanks are 
due to our colleagues Constantin Brif, Hiroshi Ezawa, Oded Kenneth, 
Israel Klich, Koichi Nakamura, Amos Ori, Shmuel Nussinov, Lev Pitaevski, 
Guy Ramon, Giuseppe Vitiello and Joshua Zak for informative comments. 
Finally, JF would like to thank Roger Balian for a useful discussion. \\

\vskip1.0cm


\newpage
\setcounter{equation}{0}
\renewcommand{\theequation}{A.\arabic{equation}}
{\bf Appendix A: {Calculational details for Section 1}}
\vskip 5mm

In this appendix we evaluate the Casimir energy through a power series 
expansion in $t$ ($t \equiv  \pi \frac{T}{T_c}$ in the notations of Section 1). This scheme requires 
regularization for each term in the expansion. The result is that, within 
such an expansion, Kirchhoff's theorem is exact, i.e., the Casimir energy 
vanishes to all orders in $1/T$ in this high temperature expansion.
This is interpreted as implying that the classical equipartition theorem is
robust -  the classical approximation, once taken, is exact to within 
power series corrections. We hasten to add that this is yet another example 
of incorrect handling of infinities which is further discussed at the end 
of this appendix.

Let us return to the expression for the Casimir energy density at finite 
temperature, Eqs. (\ref{et}) and (\ref{ect}), which, in our dimensionless units are summarized 
by 
$(x_m = \sqrt{x^2 + m^2})$,
\begin{equation}\label{ect1}
\varepsilon_c(t) = \int\limits_{0}^{\infty}xdx 
\left[\frac{x}{2}{\rm coth}(\frac{\pi x}{2t}) + \sum_{m=1}^{\infty}x_m 
{\rm coth}(\frac{\pi x_m}{2t}) - \int\limits_{0}^{\infty}dm\;{\rm coth}
(\frac{\pi x_m}{2t})\right].
\end{equation}
We now adjunct to each of the above integrals the ``standard'' 
\cite{greiner,zuber} cutoff function, 
$$ f (x_m) = \exp\left[ -\alpha x_m \right], $$ 
assuring thereby the convergence of the integrals. (We are interested in the 
$\alpha \rightarrow 0$ limit.)

Assuming the validity of this regularization scheme, we may expand (\cite{abramowitz}, p. 42),
\begin{equation}\label{coth}
{\rm coth} z = \frac{1}{z} + \frac{z}{3} - \frac{z^3}{45} +\cdots + 
\frac{2^{2n}B_{2n}}{(2n)!}z^{2n-1}\quad\quad (|z|<\pi)
\end{equation}
($B_{2n}\ {\rm are\ Bernoulli\ numbers}$).
The evaluation of $\varepsilon_c(t)$ reduces to the evaluation of terms of the form 
\cite{opher}
\begin{equation}
g_p = \int_{0}^{\infty}xdx\ \frac{x^p}{2}f(x) + 
\sum_{m=1}^{\infty}\int_{m}^{\infty}xdx\ x^pf(x) - 
\int_{0}^{\infty}dm \int_{m}^{\infty}xdx\ x^pf(x) .
\end{equation} 
The cutoff parameter, $\alpha$, is set equal to zero at the end of 
the calculations;
the result is
$$ g_p = \lim_{\alpha \rightarrow 0} (-1)^{p+1}\frac{ d^{p+1} }{d\alpha^{p+1}}
\left[ \frac{1}{2\alpha} + \frac{1}{ \alpha^2 } 
\frac{\alpha}{ \exp(\alpha)-1 } - \frac{1}{ \alpha^2 } \right] . $$
Noting that (\cite{abramowitz}, p.1105),
$$ \frac{y}{\exp(y)-1} = \sum_{n=0}^{\infty}\frac{B_{n}y^{n}}{n!},$$
and that $B_{2n+1} = 0\,\,(n\geq 1)$, we get $g_p = 0$ for $p$ even, and for $p$ odd, 
$$ g_p = \frac{B_{p+3}}{(p+3)(p+2)} . $$
Returning to (\ref{ect1}), we see that upon substituting (\ref{coth})
only terms with $p$ even occur, i.e., 
$\varepsilon_c(T) = 0$. This should be compared with the exponential decay of 
the 
Casimir energy with temperature that is implied by (\ref{ect}), and is depicted in 
Fig. 1. This ``robustness'' of the classical (erroneous) solution, i.e., 
the nonexistence of corrections to this nil result, is
 worthy of note. 
It implies that the correct, low T expression, (\ref{ect}), obtainable from 
the correct quantum statistical mechanical approach, can't be obtained by analytic 
means (i.e., power series) from the high T expansion side.\\
The case of $p = 0$ in the above expansion will be recognized as the sum over
 the constrained number 
of modes, per unit volume, with the unconstrained number deducted therefrom. 
That the result is nil constitutes a proof, based on a particular 
regularization scheme, of the 
assertion (Section 2) that the number of modes, per unit volume, is unchanged 
upon varying the plates' separation, $d$.\\ 


\newpage
\setcounter{equation}{0}
\renewcommand{\theequation}{B.\arabic{equation}}
{\bf Appendix B : {Consistency of the plasma dielectric function as a cutoff }}
\vskip 5mm

We used in Section 3 the ``plasmon mode'' as a physical regulator (at zero and at finite temperature)
in our calculations, through the dielectric function (\ref{dielectric}) and (\ref{chi}) associated with it. 
Thus, in effect, we are thinking of the ``charge sources" $\psi_\om$ as 
``electrons" in a metallic conductor $\cb$. Since the plasmon  mode is a real physical excitation in 
dielectrics, we expect it to be a consistent regulator. It is quite instructive to show that this is 
indeed the case, especially in the finite temperature case. (For the sake of clarity, we display in this appendix $\hbar$'s and $k_B$'s explicitly in the 
appropriate places.)

As is well known, 
\beq\label{plasma}
\om_p^2 = {4\pi n e^2\over m_e}\,,
\eeq
where $n$ is the electron density. In most metals, where $n/10^{23}cm^{-3}={\cO}(1)$,
$\hbar\om_p$ is of the order of 10eV, corresponding to temperatures of the order of $10^5~{}^o K$, 
well above the melting point of any metal. It is also of the same order of magnitude of a typical Fermi 
energy 
\beq\label{fermi}
\epsilon_F = {\hbar^2\over 2m_e}\left(6\pi^2 n\right)^{2/3}
\eeq 
of the electron gas in the metal. Indeed, from (\ref{plasma}) and (\ref{fermi}) we have 
\beq\label{efomp}
{\epsilon_F\over\hbar\om_p} \simeq 3.7 \left(n\over N_{Avogadro}\right)^{1/6}\,,
\eeq
which is ${\cO}(1)$ for all metals.

The electron Fermi gas is practically degenerate even well above the melting points
of all metals, and thus 
\beq\label{degenerate}
k_BT<<\epsilon_F
\eeq 
in all our considerations. Only electrons with energies close to the Fermi energy 
of the gas contribute to transport.\footnote{Recall that  the behavior of the gas is governed by 
$\xi=\lambda^3 n\sim \left(120,000 {}^o K/T\right)^{3/2}(n/10^{23}cm^{-3})$, where 
$\lambda=\sqrt{2\pi\hbar^2/m_ek_BT}$ is the thermal wave length. For typical metals 
$n/10^{23}cm^{-3}={\cO}(1)$. Thus, for $T<<120,000 {}^o K$, we have $\xi>>1$, which means that 
the gas is in a high density and low temperature state.} 
The electron Fermi gas and the gas of photons in the cavity are in thermal equilibrium. Thus, in 
the present context, and due to (\ref{efomp}), 
\beq\label{tempineq}
\hbar\om_1\equiv 2\pi k_BT<<\hbar\om_p
\eeq
as well. The frequency cutoff normally allows many Matsubara modes. It is the physics of the electron 
Fermi gas in the metal that implies that $k_BT$ be much smaller than
$\hbar\om_p$. Happily enough we need not worry about the Fermi gas becoming appreciably non-degenerate.
Thus, in particular, we may ignore any $T$ dependence of $n$, and therefore of $\om_p$ in our analysis.  
The metal will melt at temperatures much lower than that. The classical limit is reached way below such 
temperatures. 


\newpage
\setcounter{equation}{0}
\renewcommand{\theequation}{C.\arabic{equation}}
{\bf Appendix C : {Path integral calculation of the Casimir effect due to a massless scalar field in the parallel 
plates geometry}}
\vskip 5mm

In this appendix we evaluate the Casimir energy due to a massless scalar field in the parallel plate geometry 
using the path integral formalism of Section 3 (following in part \cite{kardar, nussinov1, nussinov2}) and 
analyze the classical limit. We also discuss briefly deviations from perfect conductor behavior. 

As in the text, we take the plates to lie at the planes $z=0,d$. Thus, the surface coordinates are 
${\bf u^{(\alpha)}} = (x,y,\alpha),\quad (\alpha = 1,2)$.   
According to (\ref{mplates}) and (\ref{casenergy}) the Casimir energy is given 
by\footnote{Note the slight change of 
notations compared to Section 1, where $\ce_c$ denoted the Casimir energy density.}
\beq\label{casenplat}
{\cal E}_c =  {1\over 2}\int\limits_{-\infty}^{\infty} {d\om\over 2\pi i}\, \rmtr\log~\left[  
\left(\begin{array}{cc} 1 & 0\\ 0 &  1\end{array}\right) + 
{l\om_p\over 2 \hat h_\om}\,\left(\begin{array}{cc} 1 & e^{-d\hat h_\om}\\ e^{-d\hat h_\om} & 1
\end{array}\right)\right]\,,
\eeq
where according to (\ref{hhat}) $\hat h_\om = \sqrt{-\omega^2 - \partial_x^2 - \partial_y^2 +i\epsilon}$.

As in the finite temperature case (Section 3.2), we can split 
\beq\label{splitplate}
\hat\cm_\om = \hat\cm_{\om,d} + \hat\cm_{\om,od} = 
{l\om_p\over 2 \hat h_\om}\,\left(\begin{array}{cc} 1 & 0\\ 0 & 1
\end{array}\right) + 
{l\om_p\over 2 \hat h_\om}\,\left(\begin{array}{cc} 0 & e^{-d\hat h_\om}\\ e^{-d\hat h_\om} & 0
\end{array}\right)\,.
\eeq

Tracing over the ${\bf u^{(\alpha)}}$ and inserting a complete set of momentum states (with components parallel
to the plates) we obtain formally 
\beq\label{casenplat1}
{\cal E}_c =  {\caa\over 2i}\int\limits_{-\infty}^{\infty} {d\om d^2 p\over (2\pi)^3 }\, \log~
\left[ \left(1 + {l\om_p\over 2 h}\right)^2 - \left({l\om_p\over 2 h}\right)^2\,e^{-2d h}
\right]\,,
\eeq
where $\caa$ is the plate's area, and 
\beq\label{hp}
h = \sqrt{-\om^2 + {\bf p}^2}\,.
\eeq
We now Wick-rotate to imaginary frequency $\om\rightarrow -i\om$. Thus $h\rightarrow\sqrt{\om^2 + {\bf p}^2}$. 
(This is equivalent to the zero temperature limit of the free energy (\ref{free})). 
Thus, we find
\beq\label{casenplat2}
{{\cal E}_c\over \caa} =  {1\over 2}\int\limits_0^{\infty} {h^2\,dh \over 2\pi^2 }\, \log~
\left[ \left(1 + {l\om_p\over 2 h}\right)^2 - \left({l\om_p\over 2 h}\right)^2\,e^{-2d h}
\right]\,.
\eeq
This expression is obviously divergent. However, we can remove the divergent part by subtracting
from it (in the spirit of Section 3.2.1) the self-energy of the plates (the contribution of the diagonal
part $\hat\cm_{\om,d}$),
\beq\label{selfplates}
{{\cal E}^{self}_c\over \caa} =  {1\over 2\caa}\int\limits_{-\infty}^{\infty} {d\om\over 2\pi} \rmtr\log~\left[  
{\bf 1}_{\Si} + l\om_p\hat{\cal M}_{\om,d}\right] = 
{1\over 2}\int\limits_0^{\infty} {h^2\,dh \over 2\pi^2}\, \log~
\left(1 + {l\om_p\over 2 h}\right)^2\,  
\eeq
which is {\em independent} of $d$. 
The resulting subtracted expression, the relative Casimir energy (\ref{relenergy}) 
\beq\label{casenplat3}
{{\cal E}_c - {\cal E}^{self}_c\over \caa} \equiv {\ce_c^{rel}\over\caa}= {1\over 2} \int\limits_0^{\infty} {h^2\,dh\over 2\pi^2} \, \log~
\left[ 1 - \left({l\om_p\over l\om_p + 2 h }\right)^2\,e^{-2d h}\right]
\eeq
is manifestly convergent. 
This subtraction is of course the most natural regularization within the present 
formalism, and seems perhaps more transparent than the exponential regulator (\ref{expreg}) in Section 1.
We can integrate (\ref{casenplat3}) by parts and obtain
\beq\label{casenplat4}
{\ce_c^{rel}\over \caa} = -{1\over 96\pi^2 d^3} 
\int\limits_0^{\infty} {u^3\,du\over e^u\,\left(1 + {u\over l\om_p d}\right)^2 -1}\,\left(1 + 
{{2\over l\om_p d}\over 1+ {u\over l\om_p d}}\right)\,.
\eeq
The perfect conductor limit corresponds to $\om_p d\rightarrow\infty$. Slightly 
imperfect conductors can be treated by taking $\om_p d$ in (\ref{casenplat4}) large but finite. 
By expanding (\ref{casenplat4}) in inverse powers of $l\om_p d $, we can calculate 
corrections to the Casimir energy due to deviations from perfect conductor behavior. 
We obtain\footnote{This expansion is well behaved, 
since in effect, the integrand is expanded in powers of $(u/\om_p d)e^u/(e^u -1)$ and of 
$(u/\om_p d)^2 e^u/(e^u -1)$, which do not introduce poles at $u=0$, and due to the extra factor of 
$1/(e^u -1)$ in (\ref{casenplat5}) are also well behaved for large $u$.} 
\beqra\label{casenplat5}
{\ce_c^{rel}\over \caa} &=& -{1\over 96\pi^2 d^3} 
\int\limits_0^{\infty} {u^3\,du\over e^u -1}\,\left(1 + {2\over l\om_p d} - {2\over l\om_p d} {e^u\over e^u -1} u + \cdots\right)\nonumber\\{}\nonumber\\&=& -{\pi^2\over 1440 d^3}\,\left(1- {6\over l\om_p d} + \cdots\right)\,.
\eeqra
The leading term in the last equation,$ -{\pi^2\over 1440 d^3}$, i.e., the value of the Casimir energy 
in the perfect conductor limit, is precisely one half of the $T=0$ electromagnetic Casimir energy 
(\ref{casenphys}). Also displayed in (\ref{casenplat5}) is the leading correction to the Casimir 
energy due to imperfect conductor behavior. 

The resulting (zero temperature) Casimir force, per unit area, is thus
\beqra\label{forcec}
f_c (d,0) &=&  - {1\over\caa}{\pa \ce_c^{rel}\over\pa d} = -{1\over 2\pi^2} \int\limits_0^{\infty} {h^3\,dh\over 
\left(1+ {2h\over l\om_p}\right)^2\,e^{2dh} - 1}\nonumber\\{}\nonumber\\ &=&  -{1\over 32\pi^2 d^4 } \int\limits_0^{\infty} {u^3\,du\over ({u\over l\om_p d} + 1)^2\,e^u - 1}\nonumber\\{}\nonumber\\
&=& -{\pi^2\over 480 d^4}\,\left(1 - {8\over l\om_p d} + \cdots\right)\,.
\eeqra
Note that for any value of $l\om_p d$, the denominator in the second line in (\ref{forcec}) has a simple 
zero $u=0$ (and no additional zeros), and is moreover manifestly positive for all $u>0$. Thus, the 
force (\ref{forcec}) is attractive.  We displayed in the last line of (\ref{forcec}) the expansion of the 
Casimir force in inverse powers of $l\om_p d$. The leading term in that expansion, 
\beq\label{forceperfect}
-{\pi^2\over 480 d^4}\,,
\eeq
is the Casimir force in the perfect conductor limit, and it is precisely one half of 
the electromagnetic Casimir force (\ref{casimirforce}). The next term, is the leading correction due to 
deviations from perfect conductor behavior \cite{lifshitz}\footnote{There is a slight numerical error in 
the leading correction given by \cite{lifshitz}.} \cite{mrs, hargreaves}. 
This correction is manifestly positive and thus acts to reduce (\ref{forceperfect}), which makes sense 
physically. This expression, unlike (\ref{forceperfect}), depends on the unknown parameter $l$, of 
$\cO (1)$. Having this parameter in what is in principle a measurable quantity, is the price we have to 
pay for treating the thickened boundary $\cb$, at large $\om_p$, approximately as a ``skin layer'' of 
width $l/\om_p$ in Section 3 (see the discussion in the first paragraph of Section 3.1.2 and also the 
paragraph following (\ref{gkernel})). As a consistency check, we compare 
the leading correction, $(\pi^2/ 60 d^4)\,(1/l\om_p d)$, to an analysis of the same problem 
(for the electromagnetic field) in Section 3 of \cite{mrs}. The latter calculation is based on analysis 
of electromagnetic Green's functions, and does not make a ``skin layer'' approximation. We see 
that this leading correction is of the same form of the leading correction in Eq. (3.22) of \cite{mrs}, and 
equals one half of that correction provided $l=3/2$, which is indeed of $\cO (1)$.

Obviously, we cannot use (\ref{forcec}) to study grossly imperfect conductors, nor can we use it to 
study the behavior of the Casimir force at small plate separations $d$ that fall outside the range 
(\ref{drange}), since we derived (\ref{forcec}) within the ``skin layer'' approximation. In order to 
address these issues we will have to take the width of the plates and the three dimensional plasma of 
``electrons'' that live in them (as well as perhaps additional dispersive effects \cite{lamdispersive}) 
into account, but we will not carry this analysis here.

We now turn to finite temperature analysis and the classical limit. According to (\ref{free}) and (\ref{mplatest})
the Casimir free energy is given by 
\beq\label{freeplat}
{\cal F}_c = {1\over 2\beta} \sum_{n=-\infty}^{\infty} \rmtr\log~\left[  
\left(\begin{array}{cc} 1 & 0\\ 0 &  1\end{array}\right) + {l\om_p\over 2 \hat h_n}\,
\left(\begin{array}{cc} 1 & e^{-d \hat h_n }\\ 
e^{-d \hat h_n } & 1\end{array}\right)\right]\,,
\eeq
with $\hat h_n = \sqrt{\omega_n^2 - \partial_x^2 - \partial_y^2}$. 
We concentrate now on $\cf_n$, the contribution of $\om_n$ to (\ref{freeplat}). In a similar manner to 
our analysis at $T=0$ (see (\ref{casenplat2})), we find 
\beq\label{freeplatn}
{{\cal F}_n\over \caa} =  \int\limits_{\om_n}^{\infty} {h\,dh \over 2\pi\beta }\, \log~
\left[ \left(1 + {l\om_p\over 2 h}\right)^2 - \left({l\om_p\over 2 h}\right)^2\,e^{-2d h}
\right]\,,\quad n\neq 0
\eeq
and 
\beq\label{freeplat0}
{{\cal F}_0\over \caa} =  \int\limits_0^{\infty} {h\,dh\over 4\pi\beta }\, \log~
\left[ \left(1 + {l\om_p\over 2 h}\right)^2 - \left({l\om_p\over 2 h}\right)^2\,e^{-2d h}
\right]\,.
\eeq

These expressions are manifestly divergent. However, similarly to our $T=0$ analysis, we can remove the 
divergent parts by subtracting from (\ref{freeplatn}) and (\ref{freeplat0}) (in the spirit of Section 3.2.1) 
the self free-energy of the plates (the contribution of the diagonal
part $\hat\cm_{n,d}$),
\beqra\label{freeselfplates}
{{\cal F}_c^{self}\over \caa} &=& {1\over 2\beta} \sum_{n=-\infty}^{\infty} 
\rmtr\log~\left[ {\bf 1}_{\Si} + l\om_p\hat{\cal M}_{n,d}\right]\nonumber\\{}\nonumber\\
&=& 
\int\limits_0^{\infty} {h\,dh\over 4\pi\beta }\, \log~
\left(1 + {l\om_p\over 2 h}\right)^2
 +\sum_{n\geq 1}\,\,\,\int\limits_{\om_n}^{\infty} {h\,dh \over 2\pi\beta }\, \log~
\left(1 + {l\om_p\over 2 h}\right)^2
\eeqra
which is {\em independent} of $d$. 

The resulting subtacted expression, the relative Casimir free energy (\ref{relfree}),
\beqra\label{casfreeplat}
{{\cal F}_c - {\cal F}_c^{self}\over \caa} &=& {\cf_c^{rel}\over\caa} =  
\int\limits_0^{\infty} {h\,dh\over 4\pi\beta }\, \log~
\left[ 1 - \left({l\om_p\over l\om_p + 2 h }\right)^2\,e^{-2d h}\right]\nonumber\\{}\nonumber\\
&+& \sum_{n\geq 1}\,\,\,\int\limits_{\om_n}^{\infty} {h\,dh\over 2\pi\beta }\, \log~
\left[ 1 - \left({l\om_p\over l\om_p + 2 h }\right)^2\,e^{-2d h}\right]
\eeqra
is manifestly convergent, and actually coincides with half its electromagnetic counterpart 
(\ref{freecphys}) as $\om_p d\rightarrow\infty$. 

We now consider the classical limit. At temperatures $T$ such that $2 d \om_1> 1$, that is, for 
$T>T_c/4\pi^2$, 
we see that (\ref{casfreeplat}) is dominated by its $n=0$ term 
\beqra\label{casfreeplat1}
\left({\cf_c^{rel}\over \caa}\right)_0 = 
\int\limits_0^{\infty} {h\,dh\over 4\pi\beta }\, \log~
\left[ 1 - \left({l\om_p\over l\om_p + 2 h }\right)^2\,e^{-2d h}\right]\nonumber\\{}\nonumber\\
= -{1\over 32\pi d^2\beta} 
\int\limits_0^{\infty} {u^2\,du\over e^u\,\left(1 + {u\over l\om_p d}\right)^2 -1}\,\left(1 + 
{{2\over l\om_p d}\over 1+ {u\over l\om_p d}}\right)\,,
\eeqra
with exponentially small corrections from the other terms. 
Thus, in the classical limit we indeed obtain an asymptotic behavior of 
$\cf_c^{rel}/\caa$ consistent with (\ref{classicallimit}). 
As in the $T=0$ case, we expand (\ref{casfreeplat1}) in inverse powers of $l\om_p d $:
\beqra\label{casfreeplat2}
\left({\cf_c^{rel}\over \caa}\right)_0 &=&  -{1\over 32\pi d^2\beta} 
\int\limits_0^{\infty} {u^2\,du\over e^u -1}\,\left(1 + {2\over l\om_p d} - {2\over l\om_p d} 
{e^u\over e^u -1} u + \cdots\right)\nonumber\\{}\nonumber\\&=& -{T\zeta (3)\over 16\pi d^2}\,\left(1- {4\over l\om_p d} + \cdots\right)\,.
\eeqra
The leading term in that expansion, $ -{T\zeta (3)\over 16\pi d^2}$, is precisely one half 
of its electromagnetic counterpart in (\ref{classicallimit}), with half the corresponding electromagnetic
Casimir entropy 
\beq\label{casentplat}
\cs_c^{rel} = {\zeta (3)\over 16\pi}{\caa\over  d^2}\,,
\eeq 
as expected. Also displayed in (\ref{casfreeplat2}) is the leading correction to the classical limit 
(relative) Casimir free energy due to imperfect conductor behavior. They act so as to diminish the Casimir entropy 
slightly.

Finally, for the Casimir force 
(per unit area) in the classical limit we find 
\beqra\label{classforce}
f_c (d,T) &=& -{T\over 16\pi d^3 } \int\limits_0^{\infty} {u^2\,du\over ({u\over l\om_p d} + 1)^2\,e^u - 1}
\nonumber\\{}\nonumber\\
&=&-{T\over 16\pi d^3 }\left[ \int\limits_0^{\infty} {u^2\,du\over e^u - 1} - {2\over\l\om_p d}\,
\int\limits_0^{\infty} {u^3 e^u\,du\over (e^u - 1)^2 } + \cdots\right]\nonumber\\{}\nonumber\\
&=& -{T\over 8\pi d^3 }\left(\zeta (3) -{6\zeta (3)\over l\om_p d} + \cdots \right)\,.
\eeqra


\newpage
\setcounter{equation}{0}
\renewcommand{\theequation}{D.\arabic{equation}}
{\bf Appendix D : {Exponential bounds on the terms in the expansion in Eq. (\ref{expand})}}
\vskip 5mm

It is straightforward to prove the bound
\beqra\label{bound2ndorder}
&&\left|\rmtr\,\left[{l\om_p\over
{\bf 1}_{\Si} + l\om_p\hat{\cal M}_{n,d}}\,\hat{\cal M}_{n,od}\,{l\om_p\over
{\bf 1}_{\Si} + l\om_p\hat{\cal M}_{n,d}}\,\hat{\cal M}_{n,od}\right]\right|\leq\nonumber\\{}\nonumber\\
&&\sum_{\alpha,\gam}\,{e^{-2\om_n d_{\alpha\gam}}\over \left(4\pi d_{\alpha\gam}\right)^2}\,
\left(\int_{\Si_\alpha} d^2u^\alpha d^2v^\alpha \sqrt{g_\alpha({\bf u^\alpha}) g_\alpha({\bf v^\alpha})}
\left|\langle {\bf u^\alpha}| {l\om_p\over {\bf 1}_\Si + l\om_p\hat{\cal M}_{n,d}}|{\bf v^\alpha}\rangle\right|\right)
\cdot\nonumber\\{}\nonumber\\&&
\left(\int_{\Si_\gam} d^2u^\gam d^2v^\gam \sqrt{g_\gam({\bf u^\gam}) g_\gam({\bf v^\gam})}
\left|\langle {\bf u^\gam}| {l\om_p\over {\bf 1}_\Si + l\om_p\hat{\cal M}_{n,d}}|{\bf v^\gam}\rangle\right|\right)
\eeqra 
on the leading contribution of $\hat\cm_{n,od}$ on the right hand side of (\ref{expand}). For $\om_n >>1$ we may estimate
the right hand side of (\ref{bound2ndorder}) further by using the asymptotic form (\ref{mkernelimit1}) for 
$\hat\cm_{n,d}$. Thus, we find in this limit that  
$$\langle {\bf u^\alpha}| {l\om_p\over 1 + l\om_p\hat{\cal M}_{n,d}}|{\bf v^\alpha}\rangle\rightarrow
\left({1\over l\om_p} + {1\over 2\om_n}\right)^{-1}\,{\del^{(2)}({\bf u^\alpha - v^\alpha})\over 
\sqrt{g_\alpha({\bf u^\alpha})}}\,,$$
and consequently we may estimate the right hand side of (\ref{bound2ndorder}) as
\beq\label{expestimate}
\left({1\over l\om_p} + {1\over 2\om_n}\right)^{-2}\,\sum_{\alpha\gamma}\,
{e^{-2\om_n d_{\alpha\gam}}\over \left(4\pi d_{\alpha\gam}\right)^2}\,\caa_\alpha\caa_\gam\,,
\eeq
where $\caa_\alpha$ is the area of $\Si_\alpha$.
In a similar manner we can show that the next (i.e., $4$th) order in $\hat\cm_{n,d}$ in (\ref{expand}) is bounded (when 
(\ref{mkernelimit1}) applies) by 
$$\left({1\over l\om_p} + {1\over 2\om_n}\right)^{-4}\,\sum_{\alpha\beta\gamma\delta}\,
{e^{-\om_n\left(d_{\alpha\beta} + d_{\beta\gam} + d_{\gam\delta} +  d_{\delta\alpha}\right)}\over \left(4\pi\right)^4
d_{\alpha\beta} d_{\beta\gam} d_{\gam\delta} d_{\delta\alpha}}\,\caa_\alpha\caa_\beta\caa_\gam\caa_\delta\,,$$
and so on.


\newpage



\begin{thebibliography}{99} 

\bibitem{casimir} H. B. G. Casimir {\it Proc. K. Ned. Akad. Wet.} {\bf 51} (1948) 793.

\bibitem{cas50} M. Bordag (Ed.), {\it The Casimir Effect 50 Years Later}, 
Proceedings of the 4th 
Workshop on Quantum Field Theory under the Influence of External Conditions, Leipzig, September 1998 
(World Scientific, Singapore, 1999).

\bibitem{recently}
The following list of references of recent years demonstrates the versatile applicability of the 
Casimir effect and its cousins in various areas of physics. This list is somewhat sporadic and 
by no means complete. Nevertheless, we hope it may be useful for some readers. We apologize in advance to 
authors offended due to omission of their works. Additional recent papers are listed 
in other bibliographical items below.\\In particular, note the recent extensive bibliography, containing
hundreds of references, in the resource letter \cite{lambib}.\\ 
{\em on the relevance of the Casimir effect to sonoluminescence}:\\
J. Schwinger, {\it Proc. Natl. Acad. Sci. USA} {\bf 90} (1993) 958, 2105, 4505, 7285; {\bf 91} (1994) 6473.\\  
C. Eberlein, {\it Phys. Rev. } {\bf A53} (1996) 2772, {\it Phys. Rev. Lett.} {\bf 76} (1996) 3842.\\
K. A. Milton, preprint hep-th/9811174; K. A. Milton and Y. J. Ng, {\it Phys. Rev.} {\bf E55} (1997) 4207;
{\bf E57} (1998) 55044.\\K. A. Milton, preprints hep-th/9811174 and 9811054 (published in 
\cite{cas50});\\     
I. H. Brevik, V. V. Nesterenko and I. G. Pirozhenko, {\it J. Phys.} {\bf A 31} (1998) 8661 (hep-th/9710101).\\
I. H. Brevik, V. N. Marachevsky and K. A. Milton, preprint hep-th/9810062.\\
C. E. Carlson,  C. Molina-Par\'is, J. P\'erez-Mercader and M. Visser {\it Phys. Rev.} {\bf D 56} (1997) 1262;
C. Molina-Par\'is and M. Visser {\it Phys. Rev.} {\bf D 56} (1997) 6629, {\it Phys.Lett.} {\bf B395} (1997) 76.\\
S. Liberati, F. Belgiorno, M. Visser and D. W. Sciama, {\it Phys. Rev. Lett.} {\bf 83} (1999) 678 
(quant-ph/9805023), preprint quant-ph/9805031; \\
A. Chodos and S. Groff, preprint hep-ph/9807512.\\
{\em other considerations of electromagnetic fluctuations in continuous media}:\\
M.-T. Jaekel and S. Reynaud, {\it J. Phys. I France} {\bf 1} (1991) 1395.\\
J. S. H{\o }ye and I. Brevik, {\it Physica } {\bf A 259} (1998) 165, preprint quant-ph/9903086;  
I. Brevik and H. Skurdal, {\it Physica } 
{\bf A 199} (1993) 412.\\
L. H. Ford, {\it Phys. Rev.} {\bf A 58} (1998) 4279.\\
{\em radiation from moving boundaries, vacuum friction} (see also \cite{kardar, dynamic}):\\
G. Calucci, {\it J. Phys.} {\bf A 25} (1992) 3873. \\
M.-T. Jaekel and S. Reynaud, {\it Phys. Lett.} {\bf A 167} (1992) 227; 
A. Lambrecht, M.-T. Jaekel and S. Reynaud, {\it Phys. Rev. Lett.} {\bf 77} (1996) 615.\\
C. K. Law, {\it Phys. Rev.} {\bf A 49} (1994) 433; {\it Phys. Rev. Lett.} {\bf 73} (1994) 1931.\\
V. V. Dodonov, {\it Phys. Lett.} {\bf A 207} (1995) 126; {\bf A 244} (1998) 517\\
V. E. Mkrtchian, {\it Phys. Lett.} {\bf A 207} (1995) 299\\
O. M\'eplan and C. Gignoux, {\it Phys. Rev. Lett.} {\bf 76} (1996) 408.\\
D. F. Maraduduin and P. A. Maia Neto, {\it Phys. Rev.} {\bf A 57} (1998) 1379;
P. A. Maia Neto and L. A. S. Machado, {\it Phys. Rev.} {\bf A 54} (1996) 3420.\\
P. A. Maia Neto and S. Reynaud, {\it Phys. Rev.} {\bf A 47} (1993) 1639.\\
L. S. Levitov, {\it Europhys. Lett.} {\bf 8} (1989) 499; 
J. B. Pendry, {\it Jour. Phys. C: Cond. Matt.} {\bf 9} (1997) 10301; \\
J. S. H{\o }ye and I. Brevik, {\it Physica } {\bf A 181} (1992) 413; {\bf A 196} (1993) 241.\\
{\em radiative corrections}:\\
M. Bordag and J. Lindig {\it Phys. Rev.} D {\bf 58} (1998) 045003; M. Bordag and K. Scharnhorst, 
{\it Phys. Rev. Lett.} {\bf 81} (1998) 3815 (hep-th/9807121).\\
X. Kong and F. Ravndal, {\it Phys. Rev. Lett.} {\bf 79} (1997) 545.
\\
{\em piecewise continuous strings}:\\
E. D'Hoker and P. Sikivie, {\it Phys. Rev. Lett.} {\bf 71} (1993) 1136;  E. D'Hoker, P. Sikivie and Y. Kanev, 
{\it Phys. Lett.} {\bf B347} (1995) 56.\\
I. Brevik, preprints hep-th/9811219 (published in \cite{cas50});
M. H. Berntsen, I. Brevik and S. D. Odintsov, {\it Ann. Phys.} {\bf 257} (1997) 84; I. Brevik and H. B. Nielsen, 
{\it Phys. Rev.} {\bf D 51} (1995) 1869, {\bf D 41} (1990) 1185;
I. Brevik, H. B. Nielsen and S. D. Odintsov, {\it Phys. Rev.} {\bf D 53} (1996) 3224;
I. Brevik and E. Elizalde, {\it Phys. Rev.} {\bf D 49} 
(1994) 5319; I. Brevik and R. Sollie, {\it J. Math. Phys.} {\bf 38} (1997) 2774; I. Brevik, A. A. Bytsenko
and H. B. Nielsen, {\it Class. Quant. Grav.} {\bf 12} (1995) 2881.\\
E. Elizalde and S. D. Odintsov, {\it Class. Quant. Grav.} {\bf 12} (1995) 2881.
\\
{\em massive fields}:\\
E. Elizalde, M. Bordag, K. Kirsten and S. Leseduarte, {\it J. Phys.} {\bf A31} (1998) 1743; {\it Phys. Rev.}
{\bf D56} (1997) 4896 (hep-th/9608071).
\\
{\em spherical and other simple geometries}:\\
K. A. Milton, {\it Phys. Rev.} {\bf E55} (1997) 4940; C. M . Bender and K. A. Milton, 
{\it Phys. Rev.} {\bf D 50} (1994) 6547. \\
E. Elizalde, M. Bordag and K. Kirsten, preprint hep-th/9707083; G. Esposito, A. Yu. Kamenshchik
and K. Kirsten, {\it Int. Jour. Mod. Phys.} {\bf A 14} (1999) 281, preprint hep-th/9802059.\\ 
M. Bordag and K. Kirsten, {\it Phys. Rev. }{\bf D 53} (1996) 5753 (hep-th/9608070), preprint hep-th/9812060; 
M. Bordag, M. Hellmund and K. Kirsten preprint hep-th/9905204; M. Bordag, K. Kirsten and D. Vassilevich, 
{\it Phys. Rev.} {\bf D 59} (1999) 085011 (hep-th/9811015); 
G. Cognola, E. Elizalde and K. Kirsten, preprint hep-th/9906228;\\
M.E. Bowers and C.R. Hagen, {\it Phys. Rev.} {\bf D 59} (1999) 025007 (hep-th/9806193).\\ 
I. Brevik and V. Marachevsky, preprint hep-th/9901086.\\
V.V. Nesterenko and I.G. Pirozhenko, preprint hep-th/9907192.\\
I. Klich, {\it Phys. Rev.} {\bf D 61} (2000) 025004 (hep-th/9908101). 
\\
{\em asymmetric and roughened geometries} (see also \cite{kardar, kardar1}):\\
I. Brevik and M. Lygren, {\it Ann. Phys.} {\bf 251} (1996) 157;\\
M. Bordag, G. L. Klimchitskaya and V. M. Mostepanenko, {\it Int. Jour. Mod. Phys.} {\bf A10} (1995) 2661; 
{\it Mod. Phys. Lett.} {\bf A9} (1994) 2515.
\\
{\em possible phenomenological applications}:\\
M. Bordag, B. Geyer, G. L. Klimchitskaya and V. M. Mostepanenko, {\it Phys. Rev.} {\bf D58} (1998) 075003;
M. Bordag, G. T. Gillies and V. M. Mostepanenko, {\it Phys. Rev.} {\bf D56} (1997) 6; M. Bordag, 
V. M. Mostepanenko and I. Yu. Sokolov, {\it Mod. Phys. Lett.} {\bf A9} (1994) 2671.
\\
{\em effect of non-trivial topology}:\\
E. Elizalde, {\it Z. Phys.} {\bf C44} (1989) 471.
K. Kirsten and E. Elizalde {\it Phys. Lett.} {\bf B365} (1996) 72. See also \cite{trunov}.\\
{\em singularities of the energy density near boundaries}:\\ 
X. Kong and F. Ravndal, preprint hep-th/9701022. See also \cite{ford} below.\\
{\em superconductivity}:\\
L. P. Pryadko, S. Kivelson and D. W. Hone, preprint cond-matt/9711129.

\bibitem{lambib} S. K. Lamoreaux, {\it Am. Jour. Phys.} {\bf 67} (1999) 850.

\bibitem{kardar} H. Li and M. Kardar, {\it Phys. Rev.} {\bf A46} (1992) 6490; {\it Phys. Rev. Lett.} {\bf 67} (1991) 3275.\\
R. Golestanian and M. Kardar, {\it Phys. Rev.} {\bf A58} (1998) 1713, 6490; {\it Phys. Rev. Lett.} {\bf 78} (1997) 3421; {\it Rev. Mod. Phys.} {\bf 71} 
(1999) 1233 (preprint cond-mat/9711071). 
(These papers and the papers in \cite{kardar1} focus on fluctuation 
induced forces in statistical 
mechanics.)
\bibitem{kardar1} R. Golestanian, M. Goulian and M. Kardar, {\it Europhys. Lett.} {\bf 33} (1996) 241; 
{\it Phys. Rev.} {\bf E54} (1996) 6725.

\bibitem{nussinov1} O. Kenneth and S. Nussinov, preprints hep-th/9802149, hep-th/9912102, hep-th/0001045
\bibitem{nussinov2} O. Kenneth and S. Nussinov, preprint hep-th/9912291


\bibitem{amj} V. Hushwater,  {\it Am. J. Phys.} {\bf 65} (1997) 381.\\
Hushwater refers to an earlier similar observation by G. Barton, in {\it Cavity Electrodynamics}, edited by P. Berman (Academic Press, 1994), p. 444.

\bibitem{opher} M. Revzen, R. Opher, M. Opher and  A. Mann, {\it Europhys Lett} {\bf 38} (1997) 245;  
{\it Jour. Phys.} {\bf A30} (1997) 7783; M. Revzen and A. Mann, preprint quant-phys/9803059.

\bibitem{ford} L. H. Ford and N. F. Svaiter, {\it Phys. Rev.} {\bf D 58} (1998) 065007. 

\bibitem{milonni} P. W. Milonni, {\it The Quantum Vacuum} ( Academic, Boston, 1993).

\bibitem{trunov} V. M. Mostepanenko and N. N. Trunov, {\it The Casimir Effect 
and its Applications} (Clarendon, Oxford, 1997).

\bibitem{levin} F. S. Levin and D. A. Micha (Ed.), {\it Long-Range Casimir
 Forces},  ( Plenum, New York, 1993).

\bibitem{krech} M. Krech, {\it The Casimir Effect in Critical Systems}
(World Scientific, Singapore, 1994).

\bibitem{miltrev} K. A. Milton, Invited Talk at the 17th Symposium on Theoretical Physics: Applied Field Theory,
Seoul, Korea, 1998. preprint hep-th/9901011.

\bibitem{greiner} G. Plunien, B. Mueller and W. Greiner, {\it Phys. Rep.} {\bf 134} (1986) 87.

\bibitem{zuber} C. Itzykson and J. -B. Zuber, {\it Quantum Field Theory}, 
(McGraw Hill, New York, 1980).

\bibitem{prl} S. K. Lamoreaux, {\it Phys. Rev. Lett.} {\bf 78} (1997) 5; U. Mohideen and A. Roy, 
{\it Phys. Rev. Lett.} {\bf 81} (1998) 4549.

\bibitem{fierz} M. Fierz, {\it Helv. Phys. Acta} {\bf 33} (1960) 855.

\bibitem{brown} L. S. Brown and G. J. Maclay,  {\it Phys. Rev.} {\bf 184} (1969) 1272. 

\bibitem{boyer} T. H. Boyer, {\it Phys. Rev} {\bf 174} (1968) 1764. 

\bibitem{schwinger} J. Schwinger, {\it Lett. Math. Phys.} {\bf 1} (1975) 43.

\bibitem{mrssphere} K. A. Milton, L. L. DeRaad and J. Schwinger, {\it Ann. Phys.} {\bf 115} (1978) 388.

\bibitem{miltonsphere}
K. A. Milton, {\it Ann. Phys.} {\bf 127} (1980) 49; {\bf 150} (1983) 432;\\
K. A. Milton, {\it Phys. Rev} {\bf D 22} (1980) 1441, ibid. 1444; {\bf D 27} (1983) 439.\\
K. A. Milton and L. L. DeRaad, {\it Ann. Phys.} {\bf 136} (1981) 229.

\bibitem{takagi} S. Tadaki and S. Takagi, {\it Prog. Theor. Phys.} {\bf 75} (1982) 262.

\bibitem{zeta} 
S. Blau, M. Visser and A. Wipf, {\it Nucl. Phys.} {\bf B310} (1988) 163.\\ 
R. Kantowski and K. A. Milton, {\it Phys. Rev.} {\bf D36} (1987) 3712.\\
E. Elizalde and A. Romeo, {\it Int. Jour. Mod. Phys.} {\bf A5} (1990) 1653; {\bf A7} (1992) 7365; 
{\it J. Math. Phys.} {\bf 30} (1989) 1133 (Err. {\bf 31} (1990) 771.)\\
D. Birmingham and S. Sen, {\it Ann. Phys.} {\bf 161} (1985) 121, {\bf 172} (1986) 451.\\
I. Brevik, A. A. Bytsenko, A.E. Goncalves and F.L. Williams, {\it J. Phys.} {\bf A31} (1998) 4437 (hep-th/9711159).
G. Lambiase, V. V. Nesterenko and M. Bordag, preprint hep-th/9812059.

\bibitem{elizeta} E. Elizalde, S. D. Odintsov, A. Romeo, A. A. Bytsenko and S. Zerbini, 
{\it Zeta Regularization Techniques with Applications}, (World Scientific, Singapore, 1994).

\bibitem{lifshitz} E. M. Lifshitz, {\it Sov. Phys. JETP} {\bf 2}, (1956) 73.

\bibitem{mrs} K. A. Milton, L. L. DeRaad and J. Schwinger, {\it Ann. Phys.} {\bf 115} (1978) 1.

\bibitem{wolfram}
J. Ambj{\o}rn and S. Wolfram, {\it Ann. Phys.} {\bf 147} (1983) 1, 33. 

\bibitem{baliandup} R. Balian and B. Duplantier, {\it Ann. Phys.} {\bf 112} (1978) 165; \\
see also R. Balian and B. Duplantier {\it Ann. Phys.} {\bf 104} (1977) 300.

\bibitem{balian}  R. Balian and C. Bloch, {\it Ann. Phys.} {\bf 60} (1970) 401 (Err. {\bf 84} (1974) 559); 
{\bf 64} (1971) 271 (Err. {\bf 84} (1974) 559); {\bf 69} (1972) 76. 

\bibitem{bordagt}  D. Robaschik, K. Scharnhorst and E. Wieczorek, 
{\it Ann. Phys.} {\bf 174} (1987) 401.

\bibitem{hays}
P. Hays, {\it Ann. Phys.} {\bf 121} (1979) 32. 

\bibitem{brevikold} I. Brevik and I. Clausen, {\it Phys. Rev.} {\bf D 39} (1989) 603; 
I. Brevik and H. Kolbenstvedt, {\it Ann. Phys.} {\bf 143} (1982) 179.

\bibitem{candelasphere} P. Candelas, {\it Ann. Phys.} {\bf 143} (1982) 241; {\bf 167} (1986) 257. 

\bibitem{bordag} M. Bordag, D. Robaschik and E. Wieczorek, {\it Ann. Phys.} 
{\bf 165} (1987) 192. 

\bibitem{dynamic} 
W. G. Unruh, {\it Phys. Rev.} {\bf D 14} (1976) 879 (Sec. 3).~ G. T. Moore, {\it J. Math. Phys.} {\bf 11} (1970) 2679.\\ 
S. A. Fulling and P. C. W. Davies, {\it Proc. R. Soc. Lond.} {\bf A 348} (1976) 393; {\it Proc. R. Soc. Lond.} 
{\bf A 356} (1977) 237.\\
P. Candelas and D. Deutsch, {\it Proc. R. Soc. Lond.} {\bf A 354} (1977) 79; {\it Phys. Rev.} {\bf D 20} (1979) 3063;\\
L. H. Ford and A. Vilenkin, {\it Phys. Rev.} {\bf D 25} (1982) 2569;\\ 
G. Barton and C. Eberlein, {\it Ann. Phys.} {\bf 227} (1993) 222; G. Barton, {\it Ann. Phys.} {\bf 245} (1996) 361.

\bibitem{revzen} M. Revzen, {\it Am. Jour. Phys.} {\bf 38} (1970) 611.

\bibitem{pippard} A. B. Pippard, {\it The Elements of Classical 
Thermodynamics} (Cambridge University Press, London, 1957).

\bibitem{peierles} R. Peierls, {\it Proceedings of the National Institute of Sciences of India} {\bf XX} (1954) 1.

\bibitem{abramowitz} I. S. Gradshteyn and I. M. Ryzhik, {\it Table of Integrals,
 Series and Products} (Academic Press, San Diego, 1994).

\bibitem{boyer2} T. H. Boyer, {\it Phys. Rev.} {\bf A11} (1975) 5, 1650.

\bibitem{anderson} P. W. Anderson, {\it Phys. Rev.} {\bf 130} (1963) 439. 

\bibitem{hargreaves} C. M. Hargreaves, {\it Proc. Kon. Ned. Akad. Wetensch.} {\bf 68B} (1965) 231. 

\bibitem{lamdispersive} S.K. Lamoreaux, {\it Phys. Rev.} {\bf A 59} (1999) R3149. 


\end{thebibliography}
\end{document}